%% file: paper.tex
\documentclass[usenatbib]{mn2e}

\usepackage{graphicx}
\usepackage{natbib}
\usepackage{color}
\usepackage{deluxetable}
\usepackage{amsmath}

\include{symbols}

\title[Comparing Infrared Star-Formation Rate Indicators with Optically-Derived Quantities]{Comparing Infrared Star-Formation Rate Indicators with Optically-Derived Quantities}

\author[J. E. Young, C. Gronwall, J. J. Salzer, J. L. Rosenberg]{J. E. Young$^{1}$, C. Gronwall$^{1,2}$, J. J. Salzer$^{3}$, J. L. Rosenberg$^{4}$
\footnotemark[1]\\
$^{1}$Department of Astronomy and Astrophysics,\\
The Pennsylvania State University, University Park, PA 16802\\
$^{2}$Institute for Gravitation and the Cosmos,\\
The Pennsylvania State University, University Park, PA 16802\\
$^{3}$Department of Astronomy,\\
Indiana University, Bloomington, IN\\
$^{4}$School of Physics, Astronomy, and Computational Science,\\
George Mason University, Fairfax, VA}

\begin{document}

\pagerange{\pageref{firstpage}--\pageref{lastpage}} \pubyear{2002}

\maketitle

\label{firstpage}

\input{abstract}
\input{background}
\input{observations}

\input{analysis}

\input{discussion}

\input{acknowledgements}

\input{table}

\clearpage
\input{figure}
\clearpage
\bibliography{bibfile}
\end{document}

%% file: symbols.tex
\definecolor{Blue}{rgb}{0.,0.,1.0}
\definecolor{Red}{rgb}{1.,0.,.0}

\def\micron{$\mu$m }
\def\LOG#1{\rm{log}\left(#1\right)}

\def\vs{{\it vs} }
\newcommand {\apgt} {\ {\raise-.5ex\hbox{$\buildrel>\over\sim$}}\ }
\newcommand {\aplt} {\ {\raise-.5ex\hbox{$\buildrel<\over\sim$}}\ }

\makeatletter

\newcommand{\Rom}[1]{\expandafter\@slowromancap\romannumeral #1@}
\makeatother

\newcommand{\ionl}[3]{[#1$\;${\scriptsize\Rom{#2}}]\relax$\lambda$#3}

\def\HII{H{\scriptsize II}}

\def\gtrsim{\mathrel{\hbox{\rlap{\hbox{\lower5pt\hbox{$\sim$}}}\hbox{$>$}}}}

%% file: abstract.tex
\begin{abstract}
We examine the UV reprocessing efficiencies of warm dust and polycyclic aromatic hydrocarbons (PAHs) through an analysis of the mid- and far-infrared surface luminosity densities of 85 nearby H$\alpha$-selected star-forming galaxies detected by the volume-limited KPNO International Spectroscopic Survey (KISS). Because H$\alpha$ selection is not biased toward continuum-bright objects, the KISS sample spans a wide range in stellar masses ($10^8$--$10^{12}\rm{M}_\odot$), as well as H$\alpha$ luminosity ($10^{39}$--$10^{43}\rm{ergs/s}$), mid-infrared 8.0\micron luminosity ($10^{41}$--$10^{44}\rm{ergs/s}$), and [Bw-R] color (-.1--2.2). We find that mid-infrared polycyclic aromatic hydrocarbon (PAH) emission in the Spitzer IRAC 8.0\micron band correlates with star formation, and that the efficiency with which galaxies reprocess UV energy into PAH emission depends on metallicity. We also find that the relationship between far-infrared luminosity in the Spitzer MIPS 24\micron band pass and H$\alpha$-measured star-formation rate varies from galaxy to galaxy within our sample; we do not observe a metallicity dependence in this relationship. We use optical colors and established mass-to-light relationships to determine stellar masses for the KISS galaxies; we compare these masses to those of nearby galaxies as a confirmation that the volume-limited nature of KISS avoids strong biases.  We also examine the relationship between IRAC 3.6\micron luminosity and galaxy stellar mass, and find a color-dependent correlation between the two.

\end{abstract}

%% file: background.tex
\section{Introduction}

Many lines of evidence now link mid- and far-infrared emission to star-forming activity.  For example, observations of star-forming regions in nearby galaxies \citep[e.g.,][]{Calzetti2005,PerezGonzalez} show that much of the ultraviolet emission from young stars is absorbed by dust and then re-emitted in the mid- and far-infrared. In particular, because the 7.7\micron polycyclic aromatic hydrocarbon (PAH) vibrational band falls within the Spitzer 8.0\micron band pass, ultraviolet-excited IRAC 8.0\micron PAH emission is believed to track star-formation rate (SFR) \citep[e.g.,][]{Wu,Calzetti2005}. Moreover, \cite{Wu}, \cite{Hong}, and \cite{Rosenberg} find that the luminosities of star-forming galaxies in the Spitzer 8.0\micron band correlate with SFRs measured through other means.

Likewise, thermal emission from warm dust is also tied to star-forming activity because the thermal energy budget of warm dust is dominated by heating from ultraviolet light.  Specifically, many works \citep[e.g.,][]{Wu,AlonsoHerrero,PerezGonzalez,Relano,Calzetti2010} report that luminosity in the Spitzer 24\micron band tracks SFR since this bandpass is dominated by the thermal spectrum of warm dust. Supporting this correlation, \cite{Bendo2010} find that warm dust, as measured by Spitzer MIPS 70\micron observations, traces out star-forming regions in M81, as opposed to cold dust, observed in Herschel 70-−500\micron data, whose thermal budget is dominated by evolved stars and is thus detected throughout M81.

However, modern infrared observations indicate that neither 8.0\micron nor 24\micron emission is a straightforward SFR indicator. Both of these indicators trace obscured star formation; although the majority of star formation takes place deep in heavily obscured molecular clouds, unobscured ultraviolet light is not represented in the infrared energy budget.  For this reason, any galaxy-wide feature which diminishes the effective covering factor will affect the correlation of these infrared luminosities with star-formation rate.

For example, \cite{Calzetti2007} and \cite{Engelbracht} note that 8.0$\mu$m luminosity is sensitive to both star-formation history and metallicity since PAH molecules contain carbon. Complicating matters further, the ultraviolet light that causes a relationship between PAH emission and SFR also destroys PAHs. \cite{Helou} and \cite{Bendo} note that PAH emission is strongest on the rims of \HII{} regions, and speculate that the radiation environment in the centers of star-forming complexes is simply too harsh for PAH molecules to survive.  Likewise, silicate dust, the most probable source of far-IR thermal emission, is also sensitive to metallicity, and even silicate dust can be destroyed by sufficiently high temperatures. Additionally, very small grains are subject to stochastic heating by UV photons, and may be out of equilibrium.  Complicating matters further, at low star-formation rates the energy contribution from evolved stars can become a significant part of the thermal budget of even warm dust \citep{Boquien2011}. Finally, thermal dust emission alone is ambiguous because dust-enshrouded AGN can also heat interstellar dust \citep[e.g.,][]{SandersMirabel}.

In an effort to relate existing metrics of star-formation rate and characterize the infrared energy budgets of star-forming galaxies, we present an analysis of mid-infrared (IRAC), far-infrared (MIPS), and optical photometry of 85 low-redshift H$\alpha$-bright star-forming galaxies detected by the KPNO International Spectroscopic Survey (KISS) \citep[KISS;][]{KISSI}, a quasi-volume-limited H$\alpha$ selected survey. Because H$\alpha$ selection is not biased toward continuum-bright objects, the KISS sample effectively characterizes star-forming galaxies by spanning a wide range of properties, such as stellar masses ($10^8$--$10^{12}\rm{M}_\odot$) and $\rm log(O/H)+12$ metallicities (7.8--9.2).  

While photometry in all four IRAC bands and the MIPS 24$\mu$m band is tabulated in Table \ref{tbl:photometry}, we focus on IRAC 3.6$\mu$m, IRAC 8.0$\mu$m, and MIPS 24$\mu$m because of their relevance to star formation studies. Luminosity in the IRAC 8.0$\mu$m band comes from a combination of sources \citep[e.g.,][]{Wu}. The primary contributors in the IRAC 8.0$\mu$m band are vibrational emission lines of PAHs and the thermal tail of the old stellar population. Luminosity in the MIPS 24$\mu$m band, on the other hand, samples only thermal dust \citep[e.g.,][]{Wu}. The IRAC 3.6$\mu$m band is believed to sample primarily the thermal tail of the old stellar population. By studying these fluxes in tandem, we distinguish these different sources of luminosity. The IRAC 4.5\micron and 5.8\micron bands are less useful for this particular purpose because they contain weaker PAH bands and more contamination from the stellar continuum than the 8.0\micron band.

While similar research has been conducted previously \citep[e.g.,][]{Rosenberg,Calzetti2007,Calzetti2010,Wu,AlonsoHerrero,PerezGonzalez}, KISS has the significant advantage that it evenly samples galaxies across four orders of magnitude in luminosity with observations that span a factor of 60 in wavelength range. Moreover, unlike other studies sampling small regions within galaxies \citep{Calzetti2005,PerezGonzalez}, the objective-prism spectra ensure that KISS SFRs are global SFRs and can be directly compared to integrated broadband SFR indicator candidates, such as IRAC 8.0$\mu$m and MIPS 24$\mu$m luminosities.

In Section 2, we describe the observations, basic data processing, and photometry procedures. In Section 3, we describe the analysis used to examine the connections and correlations between H$\alpha$ measured SFRs, infrared luminosities, and galaxian stellar masses. In Section 4, we discuss the implications of these relations and compare our finding to those published in other works.  For this work, we assume a $\Lambda$CDM cosmology with $H_0 = 67.04\;{\rm km\;s^{-1}\;Mpc^{-1}}$, $\Omega_m = 0.3183$, and $\Omega_\Lambda = 0.6817$ \citep{PlanckCosmology}.

%% file: observations.tex
\section{Observations}
\label{sec:observations}

\subsection{KISS H$\alpha$-Selected Sample}

The KISS field used in this work is located in the NOAO Deep Wide Field Survey \citep[NDWFS;][]{Jannuzi} in Bo\"{o}tes \citep{Jangren}. The KISS objects were selected from objective-prism spectra to have redshifts $<$ 0.095 and H$\alpha$ emission 5$\sigma$ above their median spectral continua. Objective-prism spectra are preferable to slit spectra for estimating total H$\alpha$ luminosity for extended objects because they capture all of the flux from the object rather than the small fraction that would fall on a slit. A redshift (volume) limit rather than a magnitude limit ensures that the sample is not biased toward galaxies that are intrinsically bright in the continuum.

In the KISS Bo\"{o}tes field, 131 H$\alpha$ emission line objects were identified. The initial detections were followed up with higher resolution spectroscopy using the Hobby Eberly Telescope (HET), the KPNO 2.1 meter, the MDM 2.4 meter, and the Lick 3 meter telescopes \citep{Wegner,KISSG,Melbourne,KISSV,Salzer}. Using the higher resolution spectra, 28 AGN and LINER galaxies were rejected through an extinction-corrected \ionl{N}{2}{6583}/H$\alpha$ versus \ionl{O}{3}{5007}/H$\beta$ line diagnostic diagram \citep{BPT}. Even though AGN hosts may be sites of star-forming activity, we are unable to disentangle AGN H$\alpha$ emission from that of star-forming activity, making the SFR measurement from such galaxies unreliable. We concentrated on the 98 remaining objects for our study. Metal abundance estimates were computed for all galaxies using emission-line ratios of strong lines by employing the coarse abundance method described in \cite{MelbourneSalzer} and \cite{Salzer_Metal}; typical uncertainties in  $\rm log(O/H)+12$ are 0.12 dex.

The H$\alpha$ luminosities referenced in this paper shall hereafter refer to H$\alpha$ luminosities that are measured from objective-prism data and are extinction-corrected using Balmer decrements ($\rm{c}_{\rm{H}\beta}$) from follow-up spectra.  This correction assumes a single screen model for dust extinction; in this assumption we incur an uncertainty because the Balmer decrement measurements from the follow-up spectra represent the extinction averaged over only the regions encompassed by the slit used in the follow-up spectra. Thus, this uncertainty has a systematic component in that, for objects of large angular size, where only a smaller fraction of the object was encompassed in the slit, the correction applied is more likely to deviate from the true value.  

We incur another uncertainty from the assumption of a flat stellar continuum underneath the H$\alpha$ and H$\beta$ emission lines.  With high resolution spectra, it is sometimes possible to estimate the stellar absorption and appropriately correct the emission lines; the follow-up KISS spectra are not of sufficient resolution.  The primary effect is that the computed emission-line strengths are lower than the true values, affecting the Balmer decrements.  This issue, and the ramifications of using the method we describe here, are thoroughly discussed by \cite{Brinchmann}. They report that our method can underestimate star-formation rates; for the median stellar mass of our objects, this underestimation is 12\%, and is much less, even zero, for low mass or low metallicity objects.

Normalizing to the angular sizes of the KISS galaxies in the IRAC 3.6\micron channel (see below), we list the H$\alpha$ surface luminosity densities and SFR surface densities in Table \ref{tbl:photometry}.

\
\subsection{H$\alpha$ Star-Formation Rates}

To calibrate mid- and far-infrared broadband photometry as star-formation rate indicators, we use extinction-corrected H$\alpha$ luminosity as a benchmark indicator. Because the stars contributing with any significance to the ionizing flux have short lifetimes ($<$ 20 Myr), extinction-corrected H$\alpha$ luminosity is an indicator of current star formation and is relatively independent of star-formation history. \cite{Kennicutt2012} provides an H$\alpha$ luminosity to SFR calibration using solar abundances and a Salpeter IMF (0.1 - 100 M$_\odot$):

\begin{eqnarray}
\rm{SFR}\left[\rm{M}_\odot/yr\right] = 5.4\times10^{-42}\rm{L}_{\rm{H}\alpha}\left[\rm{erg}/\rm{s}\right]\\
\rm{log}\left(\rm{SFR}\left[\rm{M}_\odot/yr\right]\right) = -7.685+\rm{log}\left(\rm{L}_{\rm{H}\alpha}\left[L_\odot\right]\right)
\end{eqnarray}

\subsection{Spitzer}

The Spitzer data used in this project came from the Spitzer Deep Wide-Field Survey \citep{SDWFS} and The Spitzer Program to Observe the NDWFS Field in Bo\"{o}tes (Soifer et al. 2004).  Spitzer IRAC\footnotemark[6] mosaics from the Spitzer Deep Wide-Field Survey cover 88 of the 98 KISS star-forming galaxies; we detect all 88 of these galaxies in IRAC 3.6\micron and 85 in IRAC 8.0\micron. Because our photometric uncertainties are dominated by the uncertainties in the absolute calibration of the IRAC instrument, around 3\%\footnotemark[6], we adopt a 3\% uncertainty for all our IRAC photometric measurements.

\footnotetext[6]{IRAC Data Handbook: http://ssc.spitzer.caltech.edu/irac/dh/}

The archival MIPS data from Soifer et al. (2004) provide this project with archival Spitzer MIPS\footnotemark[7] 24$\mu$m band coverage for 81 KISS star-forming galaxies; 74 are detected. The uncertainties in the absolute photometric calibration of the MIPS instrument are around 2\%\footnotemark[7], and we adopt a 2\% uncertainty for our MIPS photometric measurements as well. Although MIPS 70$\mu$m and MIPS 160$\mu$m imaging exists for the KISS galaxies, the object-to-noise contrast is such that the detection rate is less than fifty percent in these bands.  As a result, discussion of MIPS 70$\mu$m and MIPS 160$\mu$m photometry is not included in this work.

\footnotetext[7]{MIPS Data Handbook: http://ssc.spitzer.caltech.edu/mips/dh/}

The Spitzer Deep Wide-Field Survey provides IRAC mosaics that are calibrated, co-added, and science ready.  For the MIPS 24$\mu$m photometry we used archival Post Basic Calibrated Data mosaics. Typically, 2-4 Spitzer MIPS observations exist for each object, although a small number have only one observation. We co-added the MIPS observations to construct postage-stamp images after aligning them with centroiding software from the NOAO Image Reduction and Analysis Facility (IRAF).

\subsection{NDWFS Optical Imaging}\label{optical}

Because the KISS Bo\"{o}tes field was chosen to overlap with the heavily studied NDWFS Bo\"{o}tes field \citep{Jannuzi}, deep optical images exist for the majority of the KISS galaxies in the Bo\"{o}tes field.  In particular, this study makes use of Bw, R, and I band data for 76 of the 88 star-forming galaxies KISS galaxies covered by the Spitzer Deep Wide Field Survey to calculate stellar masses for these galaxies using well-known mass-to-light relations \citep{Bell} (see Section \ref{sec:masses}). 

As with the objective-prism H$\alpha$ measurements, we corrected the Bw, R, and I band measurements for extinction using a single-screen model along with the Balmer decrements ($\rm{c}_{\rm{H}\beta}$) from follow-up spectra.  All references to optical photometry hereafter, including the stellar masses, shall refer to the extinction corrected values.

\subsection{Detection and Photometry}\label{optical}

We calculated luminosity surface densities for each object in each of the NDWFS, Spitzer IRAC, and Spitzer MIPS images discussed in the previous two subsections (when available) using extended source detection an measurement for each object performed with SExtractor \citep{SExtractor}.  We used the IRAC 3.6\micron images as SExtractor detection images; the elliptical photometric regions computed by SExtractor based on the IRAC 3.6\micron images were used for photometry in images from all of the other bands, guaranteeing that the fluxes calculated in every band are from the same physical regions within the galaxies. In this fashion we produced extended-source photometric measurements for each object in up to eight bands:  the NDWFS Bw, R, and I bands, all four Spitzer IRAC channels, and the MIPS 24\micron channel.  The NDWFS optical images were binned and the MIPS 24\micron images were up-sampled in a flux-conserving manner to match the plate Spitzer Deep Wide Field Survey plate scale.  In all cases, the SExtractor FLUX\_AUTO was used.  We then converted these fluxes into luminosity surface densities (${\rm \Sigma_\lambda}$) using the angular area of each object in the detection image.

After detecting and measuring the KISS objects within these images, we applied aperture corrections to our IRAC and MIPS measurements using aperture corrections derived in the SWIRE data release \citep{SWIRE}, as suggested by the Spitzer IRAC Handbook\footnotemark[6]. The goals of SWIRE are similar to ours in that they rely in a comparison between the four IRAC and the MIPS 24\micron channels, and the aperture corrections that they derive are consistent with those derived by the IRAC and MIPS instrument teams. As a check, we compared the IRAC magnitudes determined with this method to those listed in the catalog in \cite{SDWFS}; typically, they were within two tenths of a magnitude.

Although the IRAC 8.0\micron channel is dominated by PAH vibrational transitions, there is, however, significant contamination from the red tail of the late-type stellar continuum in the 8.0$\mu$m band pass which must removed before any PAH-based SFR metric can be assessed. Because spectral energy distribution (SED) models \citep[e.g.,][]{LiDraine} indicate that the IRAC 3.6$\mu$m band pass samples the old stellar population almost exclusively, a dust-only 8.0$\mu$m luminosity can be created by using the 3.6$\mu$m luminosity to estimate and remove the stellar contribution in the 8.0$\mu$m band. As in \cite{Helou}, we assume that the flux in the 3.6\micron band is dominated by the stellar continuum and, extrapolating this continuum, adopt a coefficient $\beta = 0.232$ to be the ratio of the stellar component of the flux density in the 8.0\micron band to the total flux density in the 3.6\micron band. This ratio is derived from the Starburst99 model \citep{Leitherer}, and as the stellar SED in the mid-infrared regime is not a strong function of stellar age, it is likely to be a reasonable approximation. This technique is also used in other work \citep[e.g.,][]{Calzetti2007,PerezGonzalez,Rosenberg,Wu}, although the adopted value of $\beta$ differs between authors, 0.22-0.29 in \citet{Calzetti2007} and 0.26 in \cite{Wu}, for instance. 

Likewise, we remove the stellar continuum from the IRAC 4.5$\mu\rm  m$, 5.8$\mu\rm  m$, and MIPS 24\micron image measurements, using the $\beta$ values of 0.596, 0.399, and 0.032, respectively \citep{Helou}.  Note that, unlike the correction made to the 8.0\micron data, these corrections have very little impact on our analysis since the $\beta$ value for MIPS 24$\mu\rm  m$ is fairly small and that our analysis does not utilize the IRAC 4.5\micron and 5.8\micron measurements, where the contamination is more significant. All references to $\Sigma_\lambda$ in any of these channels shall hereafter refer to dust-only values. Table \ref{tbl:photometry} lists luminosity surface densities for the IRAC 3.6\micron channel, as well as dust-only surface luminosities for the remaining IRAC channels and the MIPS 24\micron channel.

Because we focus on luminosity surface densities, we largely bypass the uncertainty introduced by translating flux into luminosity via the redshift.  However, it is still necessary to account for the difference between angular diameter distance and luminosity distance; to this end, we have multiplied each of the $\Sigma_\lambda$ values by $(1+z)^4$ as a final step.  All of the values listed in Table \ref{tbl:photometry} and used in our analysis have been corrected in this way.

%% file: analysis.tex
\section{Analysis}
\label{sec:analysis}

\subsection{Mid- and Far-Infrared Correlations with Total SFR}
\label{sec:direct}

Because PAH features dominate non-stellar 8.0$\mu$m flux by more than a factor of 100 above warm dust emission, and nearly a factor of 10 above the stellar continuum \citep[][see their Figures 8 and 10]{Dale,LiDraine}, several works \citep[e.g.,][]{Calzetti2007} calibrate 8.0$\mu$m luminosity as an SFR indicator.  Using extinction-corrected H$\alpha$ luminosity as a reference indicator, we test these calibrations against the KISS sample of galaxies.

To this end, Figure \ref{fig:sfr8micron} plots H$\alpha$-measured SFR surface densities against 8.0$\mu$m luminosity surface densities for the KISS galaxies that have 8.0$\mu$m measurements, along with an error-weighted linear regression determined using the method described in \cite{AkritasBershady}. The points have been color coded for $\LOG{O/H}+12$ metallicity from follow-up spectroscopy (when available); a typical uncertainty in $\LOG{O/H}+12$ is 0.12 dex, and a color key is to the right of the plot. The merit of the 8.0\micron luminosity to SFR calibrations discussed in earlier works is evident from the correlation in Figure \ref{fig:sfr8micron}. The fitted empirical relationship is shown here, with SFRs and luminosities in units of $\rm M_\odot\,yr^{-1}\,kpc^{-2}$ and  $\rm erg\,s^{-1}\,kpc^{-2}$, respectively :

\begin{equation}
\rm log\,\Sigma_{SFR} = -39.6\pm3.44 + 0.925\pm0.0829\times log\,\Sigma_{8.0\mu m}
\end{equation}

For the purpose of comparing this relationship to earlier works, we rewrite this relationship in units of $\rm M_\odot\,yr^{-1}$ and $\rm L_\odot$:

\begin{equation}
\rm log\,SFR  = -8.5\pm0.7 + 0.93\pm0.08\times  \rm log\,L_{8.0\mu m}
\end{equation}

We list the fitted coefficients for these relations in Table \ref{tbl:calibration}, along with the analogous coefficients from other work \citep{Wu,PerezGonzalez}. Different authors generally agree on the power-law coefficient for this relationship, typically giving it a value around 0.8--1.0; moreover, these relationships agree in their predicted SFR values over the range that their calibrations overlap. However, it should be noted that the samples were collected by different means. \cite{Wu} use a magnitude-limited sample favoring large galaxies. Acknowledging this, they mention that if several dwarf galaxies had been included in their fit, the slope of their relation would have been different---closer to the one found in this work. \cite{PerezGonzalez} fit their relations to \HII{} regions in M81, but PAH emission may be sensitive to filling factors and other physical effects that would make whole galaxy relations differ from relations based on individual \HII{} regions. 

The data presented here exhibit a greater scatter about the trend line than the data presented in earlier works, with a Spearman rank-order coefficient of 0.52. Because KISS is a more representative sample of galaxies, we expect greater variation from the trend line due to galaxy-to-galaxy variation in the physical processes that drive the relationships between infrared luminosities and star-formation rates.  Additionally, we also expect deviations from the trend line due to the filling factor phenomenon, as well as random variations in the systematic uncertainties in the extinction correction (see below for discussion).

Because of the known issues involved in using PAH emission as an SFR indicator, much of the focus has shifted to 24$\mu$m far-infrared thermal dust emission as an infrared SFR indicator.  Many works \citep[e.g.,][]{AlonsoHerrero,PerezGonzalez,Calzetti2005} suggest that 24\micron luminosity is linked to ionizing photons both on the galactic scale and on the scale of star-forming complexes. This band is longward of stellar emission, thus mitigating the somewhat model-dependent correction needed when examining 8.0$\mu$m luminosity.

We plot in Figure \ref{fig:sfr24micron} extinction-corrected H$\alpha$ SFR against 24$\mu$m luminosities for 74 KISS galaxies.  In contrast to earlier works focusing on star-forming clumps within galaxies \citep[e.g.,][]{PerezGonzalez,Calzetti2005}, or LIRGs rather than normal galaxies,  \citep[e.g.,][]{AlonsoHerrero} we find that the data in Figure \ref{fig:sfr24micron} show only a weak trend of increasing $\Sigma_{24\mu m}$ with $\Sigma_{\rm SFR}$, and have a relatively low Spearman rank-order coefficient of 0.37.

These data do not unambiguously indicate any breaks in the luminosity relation seen in other work, such as the diminished 8.0\micron and 24\micron emission from dwarf galaxies seen in \cite{Wu} or the enhanced 24\micron emission from high luminosity galaxies seen in \cite{Calzetti2007}. This is likely due to the greater intrinsic scatter in the KISS sample (as discussed below).  Additionally, \cite{Calzetti2010} point out that one motivation for a piecewise indicator is the Luminous Infrared Galaxy (LIRG) mode of star formation at the high luminosity end ( $\rm{L}_{24\mu\rm{m}} \gtrsim 5\times10^{43}\rm{erg/s}$); this volume-limited sample has no LIRGs and only four objects with luminosities in this range, making a piecewise relation with a break at this luminosity both unwarranted and impossible to constrain in this study.

In order to investigate the possibility that the scatter in Figure \ref{fig:sfr24micron} is driven by systematic uncertainties in the H$\alpha$ extinction correction, we compare in Figure \ref{fig:kennicutt} the extinction corrected H$\alpha$ measurements to $\rm L_{{\rm H}\alpha(obs)}+0.02\times L_{24\mu m}$, the linear combination of uncorrected (observed) H$\alpha$ and 24\micron luminosities indicated in \cite{Kennicutt2009}.  They find that this combination of obscured and unobscured SFR indicators best reproduces the carefully corrected H$\alpha$ values they report, making it an excellent SFR indicator in cases where extinction corrections are not available.

The cloud of points in Figure \ref{fig:kennicutt} is centered around $\rm L_{H\alpha(cor)} = L_{{\rm H}\alpha(obs)}+0.02\times L_{24\mu m}$, indicating that our extinction corrections produce a corrected H$\alpha$ value that, on average, agrees with the value predicted by the relationship in \cite{Kennicutt2009} based on our 24\micron data.  There is  $\sim$~0.5~dex of scatter (similar to that shown in Figure 11 of \cite{Kennicutt2009}) but no systematic variation with $\rm{c}_{\rm{H}\beta}$ (Spearman rank-order coefficient of -0.13), suggesting that a systematic error in extinction correction is not responsible for the scatter in Figures \ref{fig:sfr24micron} and \ref{fig:kennicutt}.

\subsection{Variation in Reprocessing Efficiency}
\label{sec:efficiency}

One possible explanation for the scatter in Figures \ref{fig:sfr8micron} and \ref{fig:sfr24micron} is a variation in the UV to IR reprocessing efficiency from galaxy-to-galaxy.  For example, one might suspect that a metallicity enhancement would increase a galaxy's IR luminosity for a given SFR since both PAHs and silicate grains require elements heavier than hydrogen.  This possibility is decidedly suggested by the metallicity gradient visible in Figure \ref{fig:sfr8micron}, where the objects above the trend line tend to have slightly lower metallicities than objects below the trend line.

To explore this idea further, we plot in Figure \ref{fig:8micronefficiency_metal} the 8.0\micron efficiency; that is, the ratio of the dust-only 8.0\micron surface luminosity density to extinction-corrected H$\alpha$ luminosity surface density, \vs{} metallicity.  The data in this plot show a positive trend, with a Spearman rank-order coefficient of 0.63, indicating that more metal rich galaxies are more efficient at reprocessing UV photons into PAH vibrational emission features.  The fitted empirical relationship is:

\begin{equation}
\rm log\,\frac{f_{8.0\mu m}}{f_{H\alpha}} = -7.8\pm0.7 + 1.08\pm0.08\times \left[log\left(O/H\right)+12\right]
\end{equation}

As mentioned above, the extinction correction applied to the H$\alpha$ measurements used in the $\Sigma_{\rm SFR}$ calculation dominates the uncertainty budget in our SFR measurements; if this uncertainty had a systematic component to it, it is possible that, since metallicity and extinction are causally linked, a systematic error the extinction correction could artificially create the trend seen in Figure \ref{fig:8micronefficiency_metal}.  To address this issue, we plot in Figure \ref{fig:8micronefficiency_extinction} the 8.0\micron efficiency (described above) against the $\rm c_{H\beta}$ Balmer decrement for each object.  We conclude from the lack of correction in Figure \ref{fig:8micronefficiency_extinction} (Spearman rank-order coefficient of -0.03) that we can reject with high confidence the possibility that the relationship between 8.0\micron efficiency and metallicity is driven by systematic errors in the H$\alpha$ extinction correction.

\cite{Engelbracht} explore the idea that metallicity affects PAH reprocessing of UV starlight with a sample of galaxies designed to span a wide range in metallicity; their conclusions are similar, indicating that low metallicity galaxies are PAH deficient, or, at least PAH inefficient.  One explanation they suggest is drawn from the idea that the ISM is enriched with oxygen before carbon-rich asymptotic giant branch stars release the carbon needed to form PAHs.  We explore this idea here by in Figure \ref{fig:8micronefficiency_color}, where we plot 8.0\micron efficiency against extinction-corrected Bw-R observed color, a crude indicator of stellar age; the data in Figure  \ref{fig:8micronefficiency_color} show no correlation, with a Spearman rank-order coefficient of 0.02.

It is important to note that the lowest metallicity galaxies in this sample are more than a dex more metal rich than the lowest in the sample presented by \cite{Engelbracht}, who interpret the link between metallicity and 8.0\micron properties as a break in infrared colors around a metallicity of $\rm log(O/H)+12 \aplt 8$.  The lack of correlation of 8.0\micron efficiency with Bw-R color in this sample is not in disagreement with the concept of carbon-rich asymptotic giant branch stars affecting the PAH content of the ISM, but it does suggest that the the link between metallicity and 8.0\micron efficiency is more complex than a simple correlation with stellar age in the range sampled by KISS and probed by Bw-R colors.

\subsection{Properties of the KISS Galaxies}
\label{sec:masses}

Without greater insight into the nature of the galaxies being studied it is unclear if the correlations presented above are physically significant across a range of redshifts. To investigate the
properties of the KISS sample, we measured the stellar masses of the KISS galaxies using well-established mass-to-light ratios \citep{Bell}.  The mass-to-light ratios in \cite{Bell} that we used are based on Sloan Digital Sky Survey photometry and population synthesis models. Specifically, we utilized parameters from Table 7 in \cite{Bell} to write the logarithmic stellar mass of a galaxy, in units of $\rm M_\odot$, as its logarithmic I band luminosity plus a linear function of [B-R] color:
\begin{equation}
{\rm log\:{\cal M} = log\:L_I -0.405 + 0.518\times[B-R]},
\end{equation}
We note here that \cite{Bell} report that their optical mass-to-light relations have an uncertainty of $\sim0.1$ dex, and that we have propagated this uncertainty forward through the equations that follow.

Figure \ref{fig:masses} represents with a black histogram the masses of the 76 KISS galaxies for which Bw, R, and I band data are available.  For comparison, we have represented with a red histogram the stellar masses of galaxies from the The GALEX Ultraviolet Atlas of Nearby Galaxies \citep{GildePaz}.  These masses were also computed from broadband photometry using mass-to-light ratios from \cite{Bell}. As can be seen in Figure \ref{fig:masses}, the distribution of masses of KISS galaxies follows closely that of the galaxies from \cite{GildePaz}.  Also included in Figure \ref{fig:masses} are several Local Group galaxies; the KISS mass distribution spans these objects, and we conclude that the survey methodology behind KISS is effective at sampling typical galaxies.

Because $\lambda$~=~3.6\micron is well into the Rayleigh-Jeans tail, even for  M stars, the mid-infrared colors of stellar populations are not a strong function of age or metallicity, and are even less affected by dust obscuration and reddening than H and K bands. For these reasons, 3.6\micron luminosity is appealing as a stellar mass indicator.  While the 3.6\micron band is contaminated by emission from hot dust and the 3.3\micron PAH emission feature, these sources contribute only 5 to 15\% of the whole-galaxy luminosity flux in this band \citep{Meidt2012}. In Figure \ref{fig:masstolight} we plot integrated 3.6\micron luminosities \vs stellar masses for the 76 galaxies for which we have NDWFS photometry.  The horizontal (stellar mass) error bars are dominated by the 0.1 dex uncertainty in the \cite{Bell} mass-to-light relationships. The relationship between the parameters in Figure \ref{fig:masstolight} is fairly scattered, with a Spearman rank-order coefficient of 0.47; moreover, it is clearly dependent on Bw-R color, and, upon close inspection, bimodal.

The reason is apparent in Figure \ref{fig:BwRhistogram}, where we see a bimodal color distribution.  Previously, we reported a much tighter relationship between the 3.6\micron luminosity and stellar mass for these same galaxies \citep{YoungThesis}, however in that analysis our photometry was performed using apertures that were designed to encompass all of the objects and avoid missing light; the elliptical regions detected by SExtractor are restricted to photometrically secure areas, which favors the centers of galaxies over the outskirts.  Thus, late-type galaxies that are bulge dominated have much redder colors in the analysis presented here. We interpret the bimodality in Figure \ref{fig:BwRhistogram} as a bimodality in the colors of galaxy centers, which is still apparent, though less obvious, in the trend presented in \cite{YoungThesis}.  The mass surface densities, computed via the \cite{Bell} relationship above and the optical luminosity surface densities, are presented in Table \ref{tbl:photometry}.

As a confirmation, we plot in Figure \ref{fig:sizes} a histogram of the effective radii of the KISS galaxies, as detected by SExtractor in the IRAC 3.6\micron images.  Here we define the effective radius to be the geometric mean of the semi-major and semi-minor axes.  The distribution in Figure \ref{fig:sizes} peaks at around 1.5 kpc; from this we conclude that the regions which are photometrically secure in the SDWFS IRAC 3.6\micron images are restricted to the central portions of the galaxies.

%% file: discussion.tex
\section{Discussion}
\label{sec:kissdiscuss}

Using luminosity surface densities we confirm the correlations directly relating 8.0\micron luminosities to star-formation rates presented in earlier work \citep[][see Table \ref{tbl:calibration}]{Wu,PerezGonzalez}.  Our method has the advantage that it normalizes out galaxy size, unlike results from studies which use whole galaxy luminosities, thereby removing the possibility that the correlations presented are simply due to larger galaxies being, on average, brighter in all bands.  Our use of $\Sigma_\lambda$ rather than ${\rm L_\lambda}$ also largely sidesteps additional uncertainty in the redshift to distance calculation (see Section \ref{sec:observations}).

We find that our data show a much larger relative scatter. Because of its H$\alpha$ selection, KISS samples a broad range of star-forming galaxies, especially when compared with the samples of earlier authors, such as \cite{Wu}, who selected bright galaxies, \cite{Calzetti2005}, \cite{Calzetti2007}, and \cite{PerezGonzalez}, who studied star-forming regions in nearby galaxies, or \cite{Rosenberg}, who studied dwarf galaxies. Although the direct relationships presented here have more scatter, they are more representative of H$\alpha$-bright galactic populations by virtue of drawing upon an H$\alpha$-selected volume-limited sample. Moreover, the extinction-corrected objective-prism H$\alpha$ measurements give this sample total galaxy-wide SFRs, while other work commonly samples only parts of the target galaxies with slit spectra.

The results here are consistent with work to date in reporting that 8.0\micron PAH emission correlates with SFR with a power-law coefficient slightly less than unity, around 0.9 in this work and in \cite{PerezGonzalez}. In line with earlier work \citep[e.g.,][]{Wu,Calzetti2007,Calzetti2010}, we speculate here that the coefficient value of less than one is driven by the destruction of PAHs by ultraviolet light. This idea is in-line with observations that PAH emission is predominantly on the rims of \HII{} regions \citep{Helou,Bendo}. The power-law coefficient of around 0.9 in the PAH to SFR relation is, then, hardly unexpected as large galaxies with vigorous star formation are more likely to have many \HII{} regions than to have one monolithic region. Multiple \HII{} regions would increase the surface area to volume ratio, and with it the PAH reprocessing efficiency. 

As mentioned above, earlier works \citep[e.g.,][]{Calzetti2007,Engelbracht} note that the carbon content of PAHs make 8.0$\mu$m luminosity sensitive to both star-formation history and metallicity. We speculate, as they suggest, that galaxies with higher metallicities or richer star-formation histories might be more PAH rich and have larger ultraviolet covering factors, which is in keeping with the findings in \cite{Engelbracht}, and would also explain the trend in Figure \ref{fig:8micronefficiency_metal}.  While neither this work nor \cite{Engelbracht} possess a simple metric for star-formation history, our findings strongly complement their sample of galaxies, which were selected to span an extremely wide range in metallicities.  In particular, the lack of correlation between 8.0\micron efficiency and extinction (see Figure \ref{fig:8micronefficiency_extinction}) suggests that the scatter in the correlation between $\Sigma_{8.0\mu m}$ and ${\rm \Sigma_{SFR}}$ is not driven by older galaxies being dustier in general, but rather having a higher proportion of PAHs to silicate dust.

Directly comparing our data to the relationships presented in \cite{Wu} and \cite{PerezGonzalez}, we find that the \cite{Wu} relationship under predicts SFR for a given 8.0\micron luminosity, while the \cite{PerezGonzalez} relationship over predicts SFR for a given 8.0\micron luminosity; our data and our fitted relationship are logarithmically intermediate. This is consistent with the metallicity to PAH reprocessing efficiency relationship presented in \cite{Engelbracht}, since \cite{PerezGonzalez} examined dwarf galaxies, with low metallicities, and \cite{Wu} focused on bright galaxies, which largely represent the cores of mature galaxies.  Our volume-limited sample shows a larger spread than either of these studies and falls in between.

In contrast, we find relatively little correlation between $\Sigma_{24\mu m}$ and ${\rm \Sigma_{SFR}}$.  With a Spearman rank-order coefficient of 0.37, we cannot rule out the possibility that there is a weak correlation between these parameters, however any such relationship is dominated by galaxy-to-galaxy variation within the mass, metallicity, and luminosity range that this sample covers.  Moreover, from the color coding scheme in Figure \ref{fig:sfr24micron}, it is clear that there is no particular trend with metallicity.  Again, by normalizing to galaxy size, we have removed the effect of larger galaxies being, on average, brighter in all bands.

It is important to note that context is critical to the choice and use of SFR indicators.  Our study focuses on galaxy-wide correlations, which blend together emission from a range of environments. Since warm dust far-infrared emission is typically observed deep in \HII{} regions, studies which focus on individual \HII{} regions, or LIRGS, whose luminosities are dominated by monolithic star-forming complexes, may observe correlations with 24\micron luminosity which are very strong when applied to star-forming regions within galaxies but are blended together and blurred out in whole-galaxy photometry. Our study suggests that MIPS 24\micron analogs are not effective as whole-galaxy SFR indicators for typical field galaxies.

\section{\label{sec:level1} Conclusions}

Using our H$\alpha$-selected sample of star-forming galaxies, we confirm the calibrations from earlier work \citep[e.g.,][]{Calzetti2007,Wu,AlonsoHerrero,PerezGonzalez,Relano} indicating that IRAC 8.0$\mu$m luminosity tracks SFR as measured by H$\alpha$ emission. The physical mechanisms for this indicator, ultraviolet-excited PAH vibrational line emission, is well understood.  We also support observations linking 8.0\micron emission to metallicity, with findings that more metal rich galaxies are more efficient at processing UV photons into PAH emission.  This behavior does not appear to be correlated with extinction or Bw-R color.  Conversely, we find that MIPS 24\micron warm dust thermal emission is a poor indicator of galaxy-wide star formation.

Using optical mass-to-light ratios, we find that the KISS sample of galaxies closely mimics nearby galaxies in terms of stellar mass distribution.  We find that IRAC 3.6$\mu$m luminosity tracks stellar mass, but not sufficiently well that it can be applied without a color correction.

In future work, fitting the SEDs of KISS galaxies will allow us to characterize the typical SEDs of star-forming galaxies.  Modern SED models allow the inclusion of far-infrared dust emission. Using the 3.6\micron band to anchor the stellar mass and the optical bands to calibrate the stellar age, with the KISS sample we can estimate the PAH and dust contributions to the infrared SEDs of high-redshift star-forming galaxies observable with ALMA and Herschel.

%% file: acknowledgements.tex
\section*{Acknowledgments}
This work made use of images and/or data products provided by the NOAO Deep Wide-Field Survey (Jannuzi and Dey 1999; Jannuzi et al. 2005; Dey et al. 2005), which is supported by the National Optical Astronomy Observatory (NOAO). NOAO is operated by AURA, Inc., under a cooperative agreement with the National Science Foundation.

IRAF is distributed by the National Optical Astronomy Observatories, which are operated by the Association of Universities for Research in Astronomy, Inc., under cooperative agreement with the National Science Foundation.

This research has made use of the NASA/IPAC Infrared Science Archive, operated by the Jet Propulsion Laboratory, California Institute of Technology, under contract with the National Aeronautics and Space Administration.

The Institute for Gravitation and the Cosmos is supported by the Eberly College of Science and the Office of the Senior Vice President for Research at the Pennsylvania State University.

We thank the anonymous referee for many thoughtful comments and suggestions which helped this work achieve its science goals.

We are grateful to M. L. N. Ashby for his assistance and suggestions concerning our IRAC measurements.


We are grateful to Kristin C. Peterson for editorial work in the preparation of the paper.
This work is based [in part] on observations made with the Spitzer Space Telescope, which is operated by the Jet Propulsion Laboratory, California Institute of Technology under a contract with NASA.

%% file: table.tex
\begin{table*}
\caption{Spitzer IRAC \& MIPS Photometry}
\begin{tabular}{lcccccccccccccccc}
\hline
KISS 
&z
&${\rm log\left(\Sigma_{3.6}\right)}$ \tablenotemark{a}
&${\rm log\left(\Sigma_{4.5}\right)}$ \tablenotemark{a}
&${\rm log\left(\Sigma_{5.8}\right)}$ \tablenotemark{a}
&${\rm log\left(\Sigma_{8.0}\right)}$ \tablenotemark{a}
&${\rm log\left(\Sigma_{24}\right)}$ \tablenotemark{b}
&${\rm log\left(\Sigma_{mass}\right)}$\tablenotemark{c}
&${\rm log\left(\Sigma_{SFR}\right)}$\tablenotemark{c}
\\
&
&${\rm \left(erg\ s^{-1}kpc^{-2}\right)}$
&${\rm \left(erg\ s^{-1}kpc^{-2}\right)}$
&${\rm \left(erg\ s^{-1}kpc^{-2}\right)}$
&${\rm \left(erg\ s^{-1}kpc^{-2}\right)}$
&${\rm \left(erg\ s^{-1}kpc^{-2}\right)}$
&${\rm \left(M_\odot\ kpc^{-2}\right)}$
&${\rm \left(M_\odot\ yr^{-1}\ kpc^{-2}\right)}$

\\
\hline
2288& 0.07&41.5164 $\pm$ 0.0005&40.44 $\pm$ 0.08&41.10 $\pm$ 0.02&42.05 $\pm$ 0.01& ... &8.5 $\pm$ 0.3&-1.33 $\pm$ 0.02 \\
2289& 0.07&41.5380 $\pm$ 0.0003&40.61 $\pm$ 0.06&41.14 $\pm$ 0.02&41.99 $\pm$ 0.01& ... &8.7 $\pm$ 0.3&-0.36 $\pm$ 0.02 \\
2290& 0.07&40.294 $\pm$ 0.006&39.41 $\pm$ 0.08&39.6 $\pm$ 0.1&39.93 $\pm$ 0.06& ... & ... &-1.97 $\pm$ 0.02 \\
2291& 0.07&41.5011 $\pm$ 0.0005&40.41 $\pm$ 0.08&40.95 $\pm$ 0.02&41.82 $\pm$ 0.01& ... &8.7 $\pm$ 0.3&-0.81 $\pm$ 0.02 \\
2293& 0.07&41.4631 $\pm$ 0.0004&40.57 $\pm$ 0.06&41.14 $\pm$ 0.02&42.01 $\pm$ 0.01&42.582 $\pm$ 0.008&8.5 $\pm$ 0.3&-0.81 $\pm$ 0.02 \\
2294& 0.08&41.3728 $\pm$ 0.0004&40.25 $\pm$ 0.09&40.56 $\pm$ 0.03&41.32 $\pm$ 0.01&41.193 $\pm$ 0.009&8.8 $\pm$ 0.3&-1.62 $\pm$ 0.02 \\
2295& 0.03&41.9499 $\pm$ 0.0001&40.93 $\pm$ 0.07&42.02 $\pm$ 0.01&42.54 $\pm$ 0.01&42.568 $\pm$ 0.008& ... &-0.59 $\pm$ 0.02 \\
2296& 0.006&41.9317 $\pm$ 0.0001&41.66 $\pm$ 0.02&41.86 $\pm$ 0.01&41.86 $\pm$ 0.01&41.762 $\pm$ 0.008&7.8 $\pm$ 0.3&-1.41 $\pm$ 0.02 \\
2297& 0.07&41.4846 $\pm$ 0.0005&40.2 $\pm$ 0.1&40.32 $\pm$ 0.07&40.63 $\pm$ 0.02&40.65 $\pm$ 0.01& ... &-1.45 $\pm$ 0.02 \\
2298& 0.08&41.164 $\pm$ 0.001&40.08 $\pm$ 0.09&40.80 $\pm$ 0.02&41.68 $\pm$ 0.01&41.456 $\pm$ 0.009&8.2 $\pm$ 0.3&-0.85 $\pm$ 0.02 \\
2299& 0.07&40.382 $\pm$ 0.007&39.70 $\pm$ 0.06&40.08 $\pm$ 0.07&40.80 $\pm$ 0.01&40.82 $\pm$ 0.01&8.7 $\pm$ 0.3&-1.37 $\pm$ 0.02 \\
2300& 0.03&39.2 $\pm$ 0.1& ... & ... & ... &40.59 $\pm$ 0.02& ... &-1.36 $\pm$ 0.02 \\
2301& 0.07&40.889 $\pm$ 0.001&39.85 $\pm$ 0.08&39.2 $\pm$ 0.3&40.85 $\pm$ 0.01&40.891 $\pm$ 0.009&8.3 $\pm$ 0.3&-2.57 $\pm$ 0.02 \\
2302& 0.02&40.233 $\pm$ 0.006&38.5 $\pm$ 0.5&38.7 $\pm$ 0.8&39.66 $\pm$ 0.08&40.11 $\pm$ 0.03& ... &-2.09 $\pm$ 0.02 \\
2303& 0.08&40.830 $\pm$ 0.004&39.8 $\pm$ 0.1&40.34 $\pm$ 0.06&40.97 $\pm$ 0.01&41.24 $\pm$ 0.01&8.0 $\pm$ 0.3&-1.12 $\pm$ 0.02 \\
2304& 0.07&41.132 $\pm$ 0.001&40.0 $\pm$ 0.1&40.45 $\pm$ 0.03&41.32 $\pm$ 0.01&40.94 $\pm$ 0.01& ... &-1.54 $\pm$ 0.02 \\
2305& 0.07&41.204 $\pm$ 0.001&40.13 $\pm$ 0.08&40.77 $\pm$ 0.02&41.66 $\pm$ 0.01&41.406 $\pm$ 0.008&8.5 $\pm$ 0.3&-1.21 $\pm$ 0.02 \\
2306& 0.03&41.2522 $\pm$ 0.0004&40.0 $\pm$ 0.1&41.05 $\pm$ 0.01&41.51 $\pm$ 0.01&41.510 $\pm$ 0.008&8.5 $\pm$ 0.3&-1.37 $\pm$ 0.02 \\
2307& 0.06&41.017 $\pm$ 0.001&39.97 $\pm$ 0.08&40.52 $\pm$ 0.03&41.38 $\pm$ 0.01&41.296 $\pm$ 0.009&9.2 $\pm$ 0.3&-1.21 $\pm$ 0.02 \\
2308& 0.09&41.071 $\pm$ 0.001&39.95 $\pm$ 0.09&40.44 $\pm$ 0.03&41.37 $\pm$ 0.01&41.288 $\pm$ 0.009& ... &-1.73 $\pm$ 0.02 \\
2309& 0.02&39.482 $\pm$ 0.006&38.2 $\pm$ 0.2&39.37 $\pm$ 0.03&38.5 $\pm$ 0.1&39.72 $\pm$ 0.01&6.9 $\pm$ 0.3&-3.28 $\pm$ 0.02 \\
2310& 0.01&41.6879 $\pm$ 0.0002& ... &40.31 $\pm$ 0.09&39.91 $\pm$ 0.09&40.24 $\pm$ 0.01& ... &-3.47 $\pm$ 0.02 \\
2313& 0.05&40.992 $\pm$ 0.001&39.89 $\pm$ 0.09&40.56 $\pm$ 0.02&41.31 $\pm$ 0.01&39.3 $\pm$ 0.2&8.4 $\pm$ 0.3&-1.43 $\pm$ 0.02 \\
2315& 0.05&40.748 $\pm$ 0.006&40.14 $\pm$ 0.05&39.5 $\pm$ 0.6&40.34 $\pm$ 0.06& ... &8.4 $\pm$ 0.3&-1.42 $\pm$ 0.02 \\
2316& 0.04&41.1138 $\pm$ 0.0008&39.99 $\pm$ 0.09&41.11 $\pm$ 0.01&41.64 $\pm$ 0.01&39.85 $\pm$ 0.05&8.5 $\pm$ 0.3&-1.15 $\pm$ 0.02 \\
2318& 0.07&41.042 $\pm$ 0.002&40.22 $\pm$ 0.05&40.51 $\pm$ 0.04&41.37 $\pm$ 0.01&41.400 $\pm$ 0.009& ... &-1.34 $\pm$ 0.02 \\
2319& 0.09&40.924 $\pm$ 0.002&39.87 $\pm$ 0.09&40.37 $\pm$ 0.05&41.24 $\pm$ 0.01& ... &9.2 $\pm$ 0.3&-1.61 $\pm$ 0.02 \\
2320& 0.05&40.596 $\pm$ 0.001&39.2 $\pm$ 0.1&40.12 $\pm$ 0.02&40.42 $\pm$ 0.01&40.17 $\pm$ 0.01&8.1 $\pm$ 0.3&-2.73 $\pm$ 0.02 \\
2321& 0.08&41.3517 $\pm$ 0.0005&40.2 $\pm$ 0.1&40.48 $\pm$ 0.04&41.38 $\pm$ 0.01&38.7 $\pm$ 0.8& ... &-1.92 $\pm$ 0.02 \\
2322& 0.02&40.345 $\pm$ 0.002&39.27 $\pm$ 0.09&39.53 $\pm$ 0.07&40.10 $\pm$ 0.02&39.3 $\pm$ 0.1& ... &-2.14 $\pm$ 0.02 \\
2323& 0.04&41.132 $\pm$ 0.001&40.0 $\pm$ 0.1&40.86 $\pm$ 0.02&41.42 $\pm$ 0.01&41.437 $\pm$ 0.008&9.4 $\pm$ 0.3&-1.13 $\pm$ 0.02 \\
2324& 0.08&41.3365 $\pm$ 0.0009&40.20 $\pm$ 0.09&40.92 $\pm$ 0.02&41.81 $\pm$ 0.01&41.366 $\pm$ 0.009&8.7 $\pm$ 0.3&-1.02 $\pm$ 0.02 \\
2326& 0.02&40.987 $\pm$ 0.002&39.97 $\pm$ 0.08&40.67 $\pm$ 0.03&40.98 $\pm$ 0.01&41.665 $\pm$ 0.008&8.2 $\pm$ 0.3&-0.92 $\pm$ 0.02 \\
2327& 0.08&40.800 $\pm$ 0.003&40.04 $\pm$ 0.05&40.04 $\pm$ 0.07&40.81 $\pm$ 0.01& ... &7.6 $\pm$ 0.3&-1.22 $\pm$ 0.02 \\
2330& 0.08&41.263 $\pm$ 0.001&40.21 $\pm$ 0.08&40.77 $\pm$ 0.02&41.69 $\pm$ 0.01&41.324 $\pm$ 0.009&8.6 $\pm$ 0.3&-0.88 $\pm$ 0.02 \\
2332& 0.03&40.5501 $\pm$ 0.0009&39.1 $\pm$ 0.1&40.17 $\pm$ 0.02&40.57 $\pm$ 0.01&40.21 $\pm$ 0.01&8.1 $\pm$ 0.3&-1.58 $\pm$ 0.02 \\
2333& 0.08&40.982 $\pm$ 0.001&39.8 $\pm$ 0.1&40.46 $\pm$ 0.03&41.30 $\pm$ 0.01&41.144 $\pm$ 0.009&8.4 $\pm$ 0.3&-1.01 $\pm$ 0.02 \\
2335& 0.04&41.3806 $\pm$ 0.0003&40.27 $\pm$ 0.09&41.17 $\pm$ 0.01&41.70 $\pm$ 0.01&42.127 $\pm$ 0.008&9.4 $\pm$ 0.3&-1.69 $\pm$ 0.02 \\
2336& 0.03&41.8494 $\pm$ 0.0001&40.5 $\pm$ 0.1&41.70 $\pm$ 0.01&42.06 $\pm$ 0.01&42.229 $\pm$ 0.008& ... &-2.66 $\pm$ 0.02 \\
2337& 0.04&41.0669 $\pm$ 0.0005&39.9 $\pm$ 0.1&40.84 $\pm$ 0.01&41.39 $\pm$ 0.01&41.496 $\pm$ 0.008&9.4 $\pm$ 0.3&-1.84 $\pm$ 0.02 \\

\label{tbl:photometry}
\end{tabular}
\tablenotetext{a}{Dominated by a 3\%. uncertainty in IRAC calibration.}
\tablenotetext{b}{Dominated by a 2\% uncertainty in MIPS calibration.}
\tablenotetext{c}{Dominated by a 10\% uncertainty in the mass-to-light ratios.}
\end{table*}

\begin{table*}
\begin{tabular}{lcccccccccccccccc}
\hline
KISS
&z
&${\rm log\left(\Sigma_{3.6}\right)}$ \tablenotemark{a}
&${\rm log\left(\Sigma_{4.5}\right)}$ \tablenotemark{a}
&${\rm log\left(\Sigma_{5.8}\right)}$ \tablenotemark{a}
&${\rm log\left(\Sigma_{8.0}\right)}$ \tablenotemark{a}
&${\rm log\left(\Sigma_{24}\right)}$ \tablenotemark{b}
&${\rm log\left(\Sigma_{mass}\right)}$ \tablenotemark{c}
&${\rm log\left(\Sigma_{SFR}\right)}$ \tablenotemark{c}
\\
&
&${\rm \left(erg\ s^{-1}kpc^{-2}\right)}$
&${\rm \left(erg\ s^{-1}kpc^{-2}\right)}$
&${\rm \left(erg\ s^{-1}kpc^{-2}\right)}$
&${\rm \left(erg\ s^{-1}kpc^{-2}\right)}$
&${\rm \left(erg\ s^{-1}kpc^{-2}\right)}$
&${\rm \left(M_\odot\ kpc^{-2}\right)}$
&${\rm \left(M_\odot\ yr^{-1}\ kpc^{-2}\right)}$

\\
\hline
2339& 0.05&40.936 $\pm$ 0.001&39.8 $\pm$ 0.1&40.52 $\pm$ 0.02&41.26 $\pm$ 0.01&41.044 $\pm$ 0.009&9.3 $\pm$ 0.3&-1.27 $\pm$ 0.02 \\
2341& 0.08&41.6116 $\pm$ 0.0003&41.07 $\pm$ 0.03&41.40 $\pm$ 0.01&41.99 $\pm$ 0.01&40.12 $\pm$ 0.04&9.8 $\pm$ 0.3&-1.30 $\pm$ 0.02 \\
2342& 0.08&41.4393 $\pm$ 0.0003&40.51 $\pm$ 0.06&40.77 $\pm$ 0.02&41.66 $\pm$ 0.01&41.712 $\pm$ 0.008&9.7 $\pm$ 0.3&-1.99 $\pm$ 0.02 \\
2345& 0.08&41.3249 $\pm$ 0.0006&40.21 $\pm$ 0.09&40.76 $\pm$ 0.02&41.69 $\pm$ 0.01&40.02 $\pm$ 0.04&9.7 $\pm$ 0.3&-0.89 $\pm$ 0.02 \\
2346& 0.04&39.6 $\pm$ 0.1& ... &40.2 $\pm$ 0.1& ... &39.2 $\pm$ 0.5& ... &-0.71 $\pm$ 0.02 \\
2347& 0.02&41.3352 $\pm$ 0.0004&40.2 $\pm$ 0.1&41.20 $\pm$ 0.01&41.63 $\pm$ 0.01&42.104 $\pm$ 0.008&9.3 $\pm$ 0.3&-1.14 $\pm$ 0.02 \\
2349& 0.01&41.006 $\pm$ 0.001&40.42 $\pm$ 0.03&41.02 $\pm$ 0.01&41.34 $\pm$ 0.01&42.326 $\pm$ 0.008&9.0 $\pm$ 0.3&-0.85 $\pm$ 0.02 \\
2351& 0.02&41.2552 $\pm$ 0.0002&40.11 $\pm$ 0.09&41.18 $\pm$ 0.01&41.55 $\pm$ 0.01&42.069 $\pm$ 0.008&9.4 $\pm$ 0.3&-1.37 $\pm$ 0.02 \\
2352& 0.04&42.74184 $\pm$ 4e-05&42.28 $\pm$ 0.03&42.83 $\pm$ 0.01&42.32 $\pm$ 0.01&41.573 $\pm$ 0.009& ... &-2.31 $\pm$ 0.02 \\
2353& 0.09&41.5328 $\pm$ 0.0004&40.62 $\pm$ 0.06&41.18 $\pm$ 0.02&42.11 $\pm$ 0.01&42.003 $\pm$ 0.008&9.7 $\pm$ 0.3&-0.39 $\pm$ 0.02 \\
2354& 0.09&41.077 $\pm$ 0.002&40.07 $\pm$ 0.08&40.64 $\pm$ 0.04&41.41 $\pm$ 0.01& ... &9.1 $\pm$ 0.3&-1.27 $\pm$ 0.02 \\
2355& 0.07&41.4920 $\pm$ 0.0002&40.47 $\pm$ 0.07&40.95 $\pm$ 0.02&41.80 $\pm$ 0.01&41.667 $\pm$ 0.008&9.9 $\pm$ 0.3&-0.82 $\pm$ 0.02 \\
2356& 0.06&41.4922 $\pm$ 0.0004&40.2 $\pm$ 0.1&40.96 $\pm$ 0.02&41.70 $\pm$ 0.01& ... & ... &-0.57 $\pm$ 0.02 \\
2357& 0.03&40.420 $\pm$ 0.002&39.3 $\pm$ 0.1&39.6 $\pm$ 0.1&40.12 $\pm$ 0.02&40.44 $\pm$ 0.01&8.8 $\pm$ 0.3&-2.07 $\pm$ 0.02 \\
2358& 0.05&41.3576 $\pm$ 0.0007&40.29 $\pm$ 0.08&41.17 $\pm$ 0.01&41.86 $\pm$ 0.01&38 $\pm$ 5&9.4 $\pm$ 0.3&-1.32 $\pm$ 0.02 \\
2360& 0.03&41.043 $\pm$ 0.001&39.95 $\pm$ 0.09&40.35 $\pm$ 0.03&41.13 $\pm$ 0.01&41.223 $\pm$ 0.009&9.3 $\pm$ 0.3&-1.31 $\pm$ 0.02 \\
2361& 0.03&41.4100 $\pm$ 0.0003&40.2 $\pm$ 0.1&41.18 $\pm$ 0.01&41.70 $\pm$ 0.01&41.756 $\pm$ 0.008&9.0 $\pm$ 0.3&-1.44 $\pm$ 0.02 \\
2362& 0.03&41.6709 $\pm$ 0.0001&40.4 $\pm$ 0.1&41.40 $\pm$ 0.01&41.91 $\pm$ 0.01&42.004 $\pm$ 0.008& ... &-0.61 $\pm$ 0.02 \\
2363& 0.08&41.267 $\pm$ 0.001&41.17 $\pm$ 0.02&41.44 $\pm$ 0.01&41.59 $\pm$ 0.01&42.136 $\pm$ 0.008&9.0 $\pm$ 0.3&0.26 $\pm$ 0.02 \\
2364& 0.07&41.3658 $\pm$ 0.0003&39.7 $\pm$ 0.3&39.97 $\pm$ 0.09&40.08 $\pm$ 0.04&40.11 $\pm$ 0.02&10.0 $\pm$ 0.3&-2.06 $\pm$ 0.02 \\
2366& 0.09&40.803 $\pm$ 0.002&39.7 $\pm$ 0.1&40.18 $\pm$ 0.05&41.05 $\pm$ 0.01&40.11 $\pm$ 0.04&8.9 $\pm$ 0.3&-0.61 $\pm$ 0.02 \\
2367& 0.03&40.492 $\pm$ 0.002&39.57 $\pm$ 0.07&39.87 $\pm$ 0.05&40.32 $\pm$ 0.01&40.70 $\pm$ 0.01&8.9 $\pm$ 0.3&-1.84 $\pm$ 0.02 \\
2368& 0.04&40.657 $\pm$ 0.004&40.24 $\pm$ 0.03&40.64 $\pm$ 0.02&41.04 $\pm$ 0.01&41.968 $\pm$ 0.008&8.9 $\pm$ 0.3&-1.17 $\pm$ 0.02 \\
2369& 0.04&41.1477 $\pm$ 0.0006&40.13 $\pm$ 0.07&40.55 $\pm$ 0.02&41.02 $\pm$ 0.01&41.777 $\pm$ 0.008&9.6 $\pm$ 0.3&-1.83 $\pm$ 0.02 \\
2370& 0.03&41.1367 $\pm$ 0.0006&39.9 $\pm$ 0.1&40.85 $\pm$ 0.01&41.34 $\pm$ 0.01&39.4 $\pm$ 0.1& ... &-2.27 $\pm$ 0.02 \\
2371& 0.02&41.3614 $\pm$ 0.0007&40.0 $\pm$ 0.1&39.6 $\pm$ 0.2&40.02 $\pm$ 0.07& ... & ... &-1.79 $\pm$ 0.02 \\
2372& 0.07&41.218 $\pm$ 0.001&40.0 $\pm$ 0.1&40.59 $\pm$ 0.03&41.52 $\pm$ 0.01&41.234 $\pm$ 0.009&9.6 $\pm$ 0.3&-1.19 $\pm$ 0.02 \\
2373& 0.06&41.294 $\pm$ 0.001&40.1 $\pm$ 0.1&40.80 $\pm$ 0.02&41.65 $\pm$ 0.01&41.456 $\pm$ 0.009&9.7 $\pm$ 0.3&-0.63 $\pm$ 0.02 \\
2374& 0.08&40.952 $\pm$ 0.002&39.92 $\pm$ 0.08&40.42 $\pm$ 0.04&41.16 $\pm$ 0.01& ... &9.0 $\pm$ 0.3&-1.12 $\pm$ 0.02 \\
2376& 0.04&41.3520 $\pm$ 0.0002&40.0 $\pm$ 0.1&40.46 $\pm$ 0.03&40.70 $\pm$ 0.01&41.092 $\pm$ 0.008&9.6 $\pm$ 0.3&-1.68 $\pm$ 0.02 \\
2377& 0.08&40.991 $\pm$ 0.001&39.93 $\pm$ 0.08&40.29 $\pm$ 0.04&41.25 $\pm$ 0.01&41.100 $\pm$ 0.009& ... &-140 $\pm$ -100 \\
2380& 0.06&41.070 $\pm$ 0.001&40.03 $\pm$ 0.08&40.70 $\pm$ 0.02&41.45 $\pm$ 0.01&41.403 $\pm$ 0.009&9.2 $\pm$ 0.3&-1.44 $\pm$ 0.02 \\
2382& 0.02&41.027 $\pm$ 0.001&39.7 $\pm$ 0.1&40.62 $\pm$ 0.02&40.98 $\pm$ 0.01&40.96 $\pm$ 0.01& ... &-1.74 $\pm$ 0.02 \\
2386& 0.03&41.1984 $\pm$ 0.0007&40.0 $\pm$ 0.1&41.03 $\pm$ 0.01&41.50 $\pm$ 0.01&41.323 $\pm$ 0.009& ... &-1.53 $\pm$ 0.02 \\
2387& 0.07&41.5607 $\pm$ 0.0003&40.51 $\pm$ 0.08&40.95 $\pm$ 0.02&41.72 $\pm$ 0.01&42.018 $\pm$ 0.008& ... &-140 $\pm$ -100 \\
2388& 0.05&41.1820 $\pm$ 0.0006&40.0 $\pm$ 0.1&40.90 $\pm$ 0.01&41.50 $\pm$ 0.01&41.478 $\pm$ 0.008&9.5 $\pm$ 0.3&-1.80 $\pm$ 0.02 \\
2389& 0.09&41.2708 $\pm$ 0.0009&40.23 $\pm$ 0.08&41.03 $\pm$ 0.02&41.91 $\pm$ 0.01&41.726 $\pm$ 0.008&9.4 $\pm$ 0.3&-1.12 $\pm$ 0.02 \\
2390& 0.03&41.92831 $\pm$ 7e-05&40.88 $\pm$ 0.08&42.04 $\pm$ 0.01&42.50 $\pm$ 0.01&42.715 $\pm$ 0.008& ... &-0.63 $\pm$ 0.02 \\
2392& 0.09&40.978 $\pm$ 0.002&39.89 $\pm$ 0.09&40.54 $\pm$ 0.03&41.28 $\pm$ 0.01&41.01 $\pm$ 0.01&9.2 $\pm$ 0.3&-0.41 $\pm$ 0.02 \\
2393& 0.03&41.2575 $\pm$ 0.0003&40.1 $\pm$ 0.1&41.04 $\pm$ 0.01&41.54 $\pm$ 0.01&41.608 $\pm$ 0.008& ... &-1.34 $\pm$ 0.02 \\
2394& 0.04&41.3486 $\pm$ 0.0005&40.23 $\pm$ 0.09&41.18 $\pm$ 0.01&41.81 $\pm$ 0.01&41.730 $\pm$ 0.008&9.4 $\pm$ 0.3&-1.00 $\pm$ 0.02 \\
2395& 0.03&41.7362 $\pm$ 0.0001&40.89 $\pm$ 0.05&41.11 $\pm$ 0.02&41.28 $\pm$ 0.01&41.975 $\pm$ 0.008& ... &-1.33 $\pm$ 0.02 \\
2397& 0.08&41.6945 $\pm$ 0.0003&40.78 $\pm$ 0.06&41.24 $\pm$ 0.02&42.13 $\pm$ 0.01&42.029 $\pm$ 0.008&9.8 $\pm$ 0.3&-0.93 $\pm$ 0.02 \\
2398& 0.03&40.821 $\pm$ 0.001&39.7 $\pm$ 0.1&41.36 $\pm$ 0.01&40.84 $\pm$ 0.01& ... & ... &-1.81 $\pm$ 0.02 \\
2399& 0.04&41.5048 $\pm$ 0.0002&40.39 $\pm$ 0.09&41.42 $\pm$ 0.01&41.87 $\pm$ 0.01&41.927 $\pm$ 0.008&9.5 $\pm$ 0.3&-1.40 $\pm$ 0.02 \\
2403& 0.04&40.713 $\pm$ 0.006& ... & ... & ... & ... &8.7 $\pm$ 0.3&-0.74 $\pm$ 0.02 \\
2406& 0.03&41.2338 $\pm$ 0.0009&39.8 $\pm$ 0.1&40.91 $\pm$ 0.02&41.24 $\pm$ 0.01&41.265 $\pm$ 0.009& ... &-1.41 $\pm$ 0.02 \\
2408& 0.08&40.915 $\pm$ 0.002&39.6 $\pm$ 0.1&40.38 $\pm$ 0.03&41.10 $\pm$ 0.01& ... & ... &-140 $\pm$ -100 \\

\end{tabular}
\tablenotetext{a}{Dominated by a 3\%. uncertainty in IRAC calibration.}
\tablenotetext{b}{Dominated by a 2\% uncertainty in MIPS calibration.}
\tablenotetext{c}{Dominated by a 10\% uncertainty in the mass-to-light ratios.}
\end{table*}

\begin{table*}
\caption{SFR Calibration Coefficients}
\begin{tabular}{lcrr}
\hline
Source
&a
&b
\\
\hline
This Work                 & -8.5$\pm$0.7    & 0.93$\pm$0.08 \\
\cite{Wu}                 & -10.03$\pm$0.03 & 1.09 $\pm$0.06   \\
\cite{PerezGonzalez}      & -7.9$\pm$0.3    & 0.87 $\pm$0.03 \\
\label{tbl:calibration}
\end{tabular}
\tablenotetext{}{ The fits are of the form $\rm{log}\ SFR\left(M_\odot\right) = a + b\times\rm{log}\ L\left(L_\odot\right)$.}
\end{table*}

%% file: figure.tex
\begin{figure}
  \includegraphics[width=\linewidth]{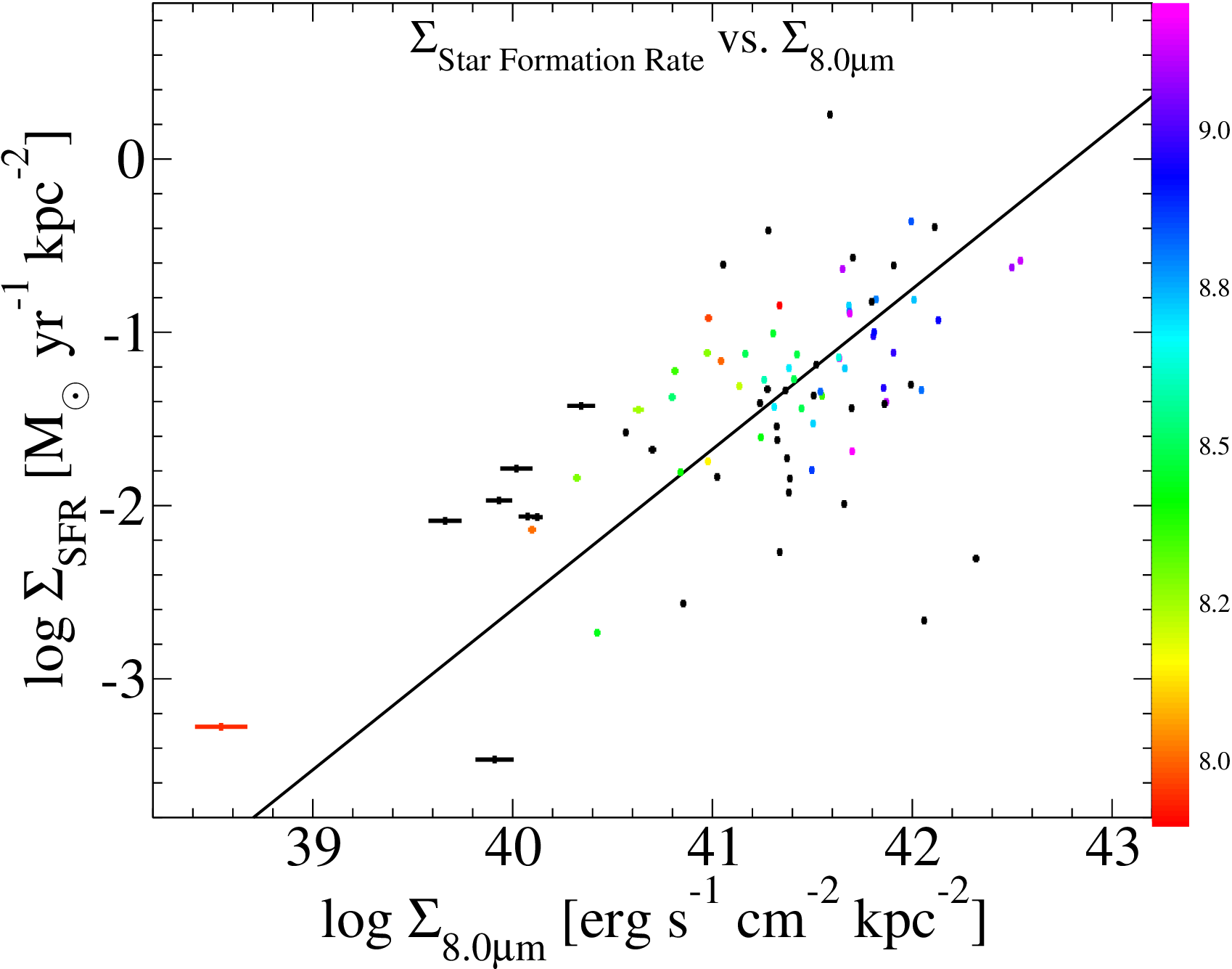}
  \caption{H$\alpha$-derived SFR surface density \vs 8.0$\mu$m luminosity surface density, with a line of best fit. Points are color coded by $\LOG{O/H}+12$ metallicity, with a color key to the right of the plot. Note that metallicity increases with 8.0\micron density and SFR density.}\label{fig:sfr8micron}
\end{figure}

\begin{figure}
  \includegraphics[width=\linewidth]{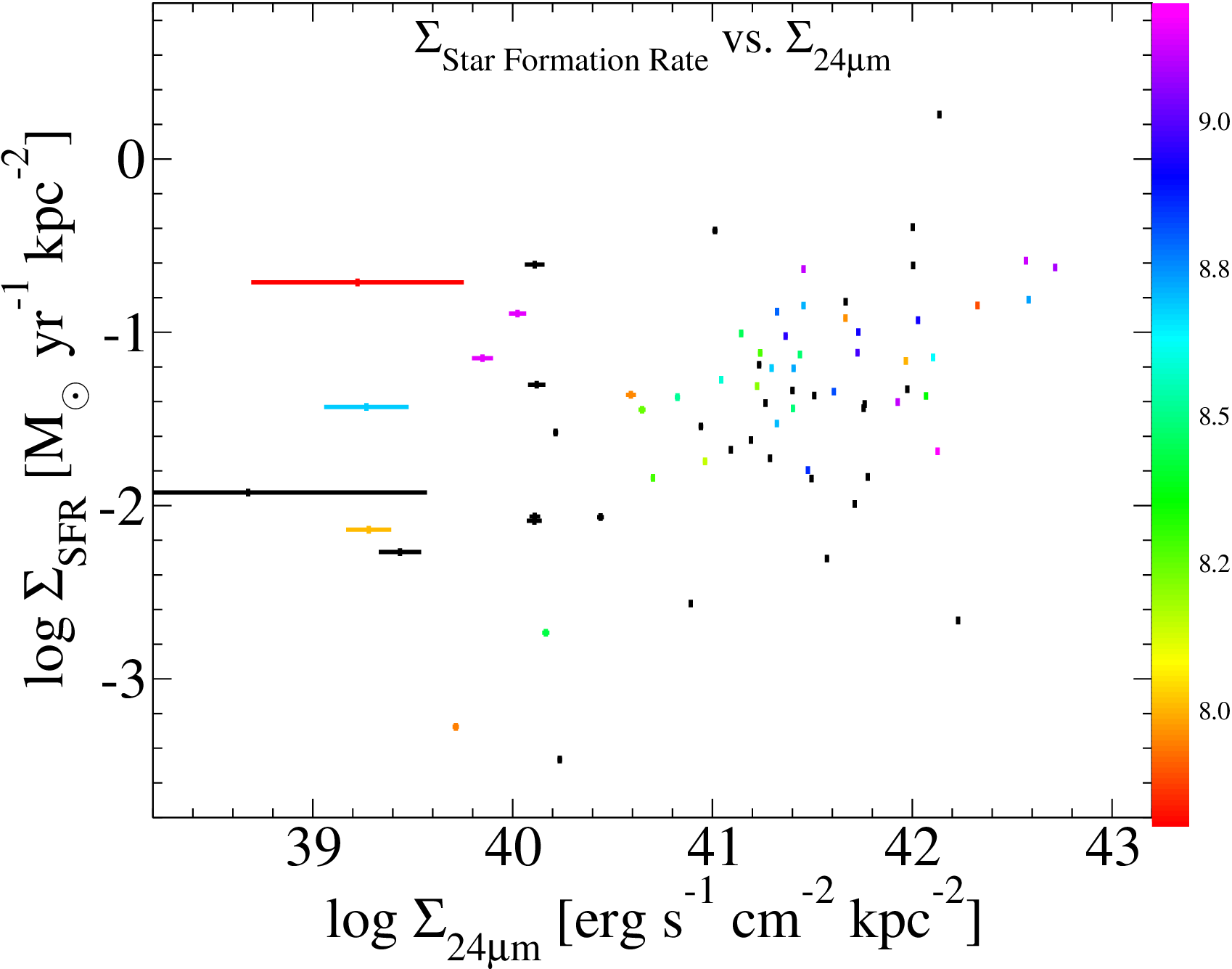}
  \caption{H$\alpha$-derived SFR surface density \vs 24$\mu$m luminosity surface density. Points are color coded by $\LOG{O/H}+12$ metallicity, with a color key to the right of the plot. Compared to Figure \ref{fig:sfr8micron}, there is relatively little trend.\label{fig:sfr24micron}}
\end{figure}

\begin{figure}
  \includegraphics[width=\linewidth]{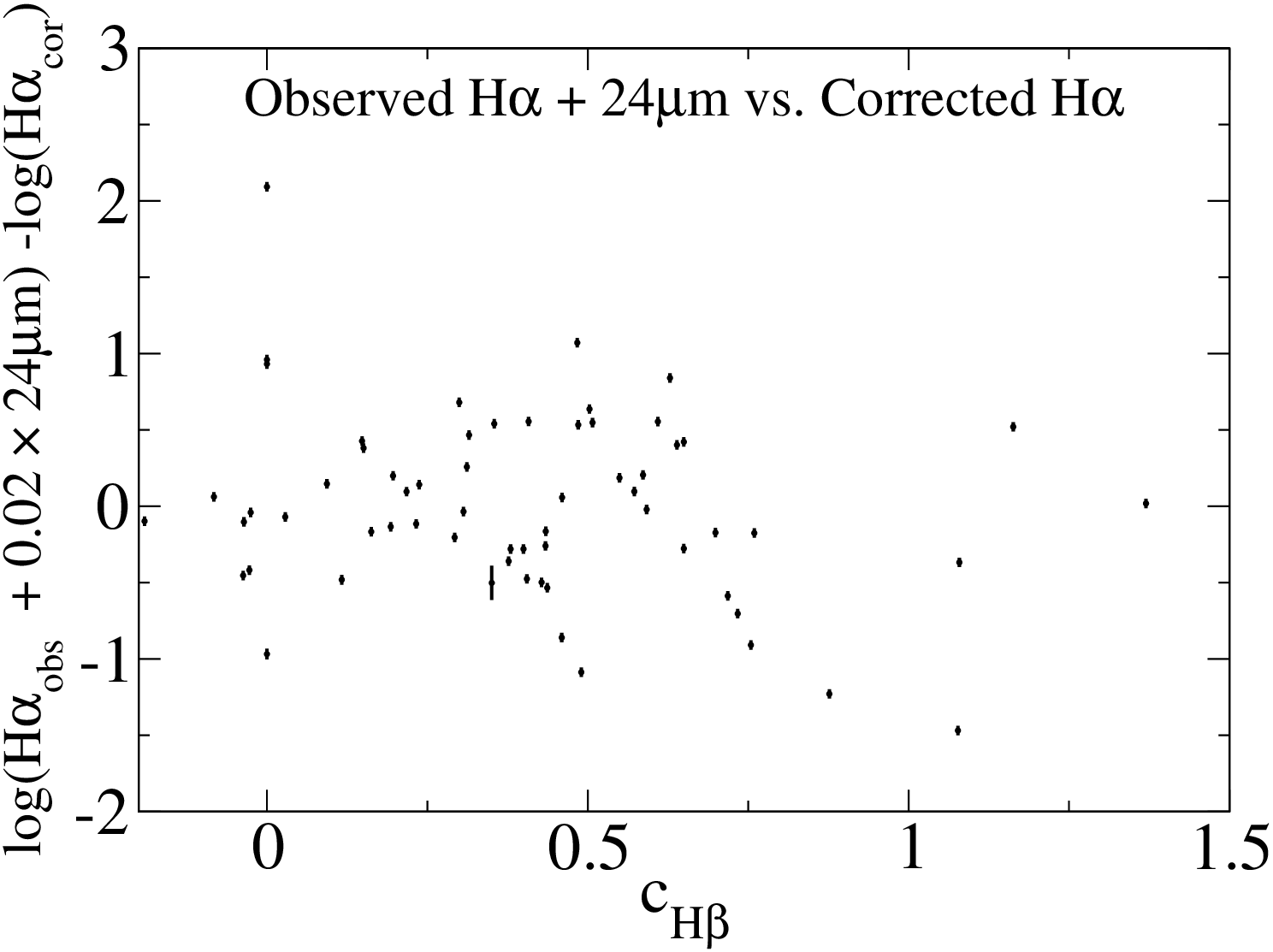}
  \caption{A comparison of a linear combination of uncorrected (observed) H$\alpha$ with 24\micron \citep{Kennicutt2009} with extinction corrected H$\alpha$, plotted against the Balmer decrement. The comparison values center around zero, and has no trend, with a Spearman rank-order coefficient of -0.13}
  \label{fig:kennicutt}
\end{figure}

\begin{figure}
  \includegraphics[width=\linewidth]{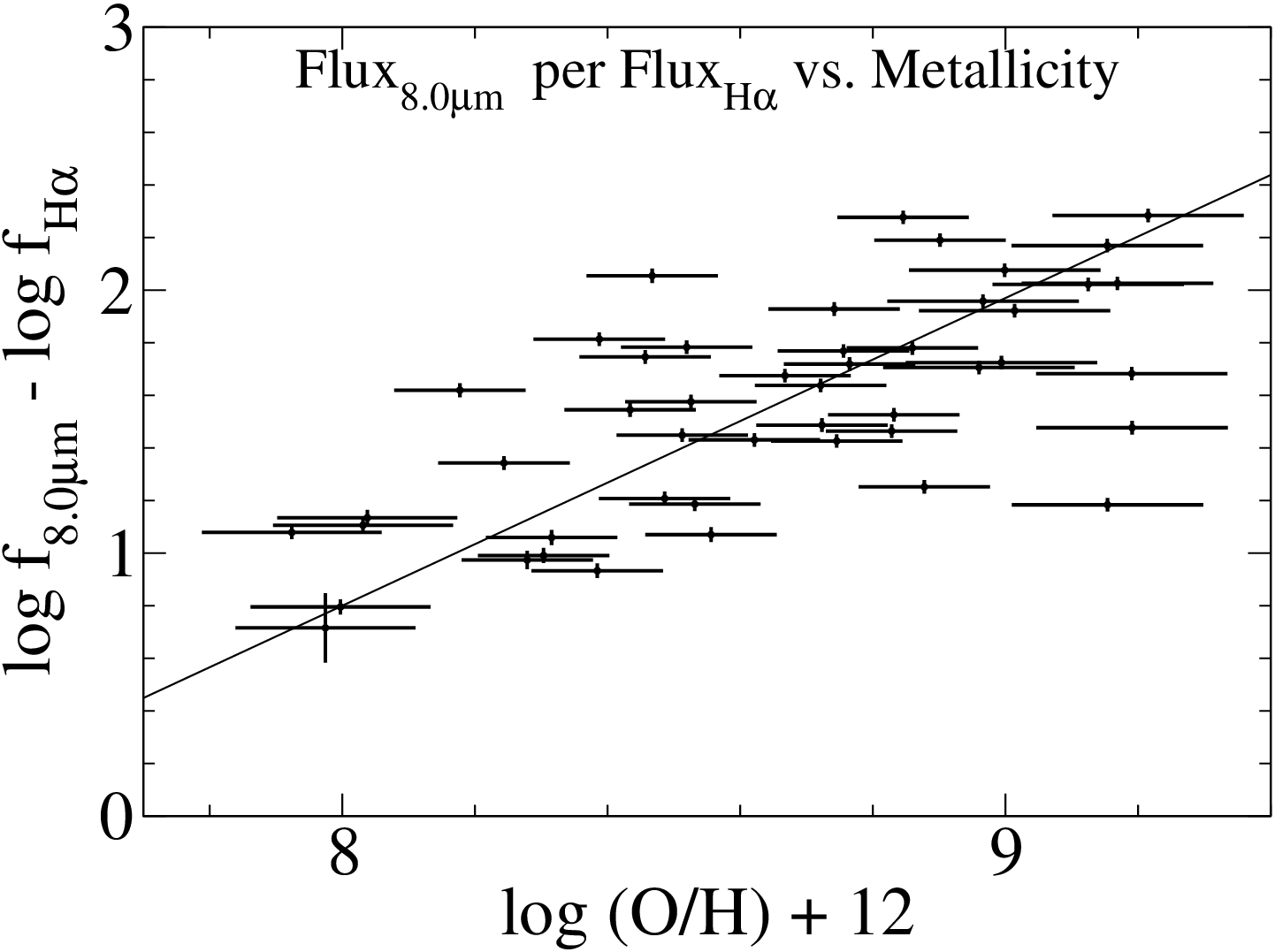}
  \caption{The ratio of 8.0\micron flux to extinction-corrected H$\alpha$ flux \vs log(O/H)+12 metallicity, along with a trend line; this figure shows a distinct trend, with a Spearman rank-order coefficient of 0.63}
  \label{fig:8micronefficiency_metal}
\end{figure}

\begin{figure}
  \includegraphics[width=\linewidth]{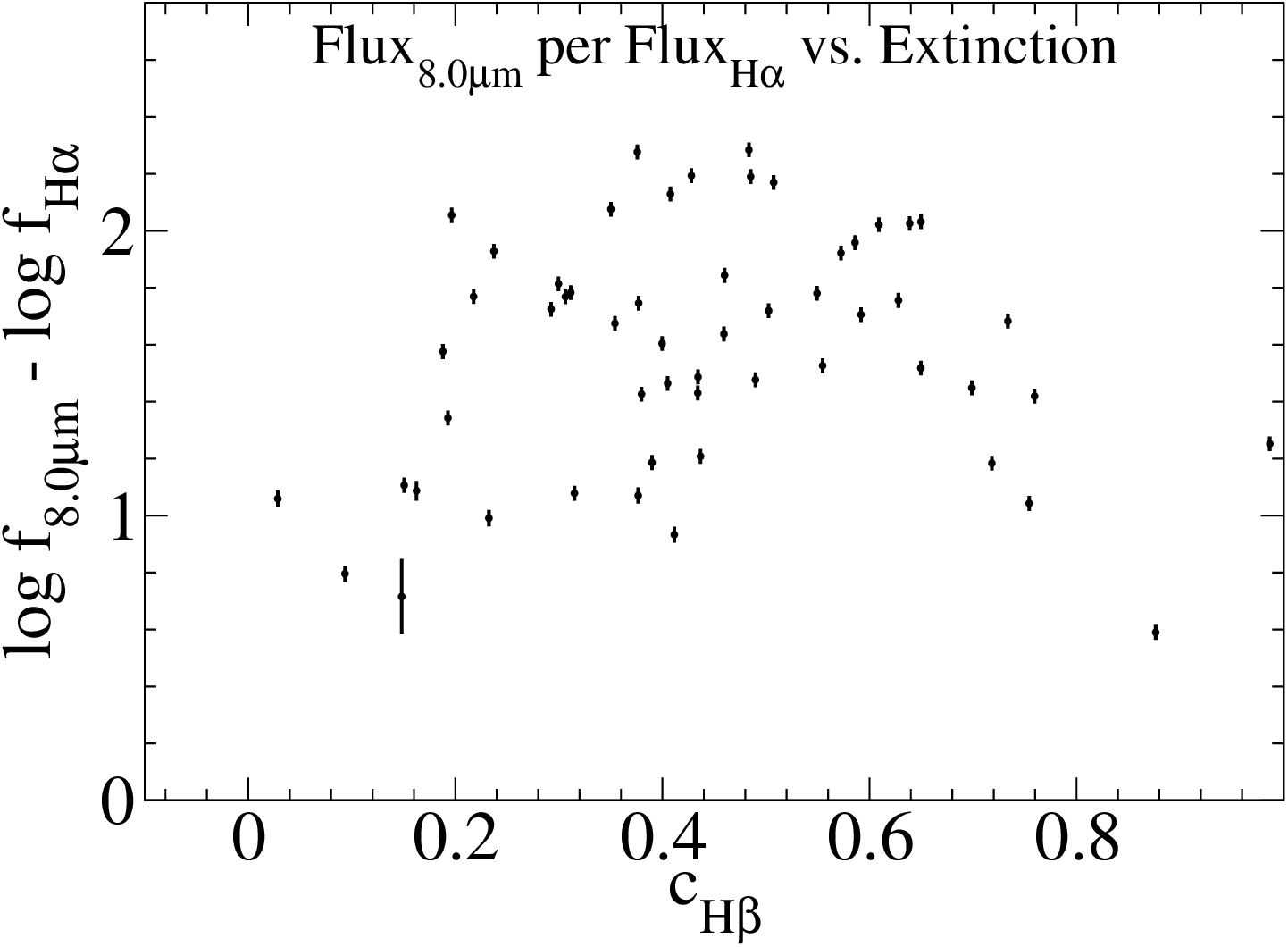}
  \caption{The ratio of 8.0\micron flux to extinction-corrected H$\alpha$ flux \vs the c$_{\rm H\beta}$ Balmer decrement; this figure shows no trend, with a Spearman rank-order coefficient of -0.02.}
  \label{fig:8micronefficiency_extinction}
\end{figure}

\begin{figure}
  \includegraphics[width=\linewidth]{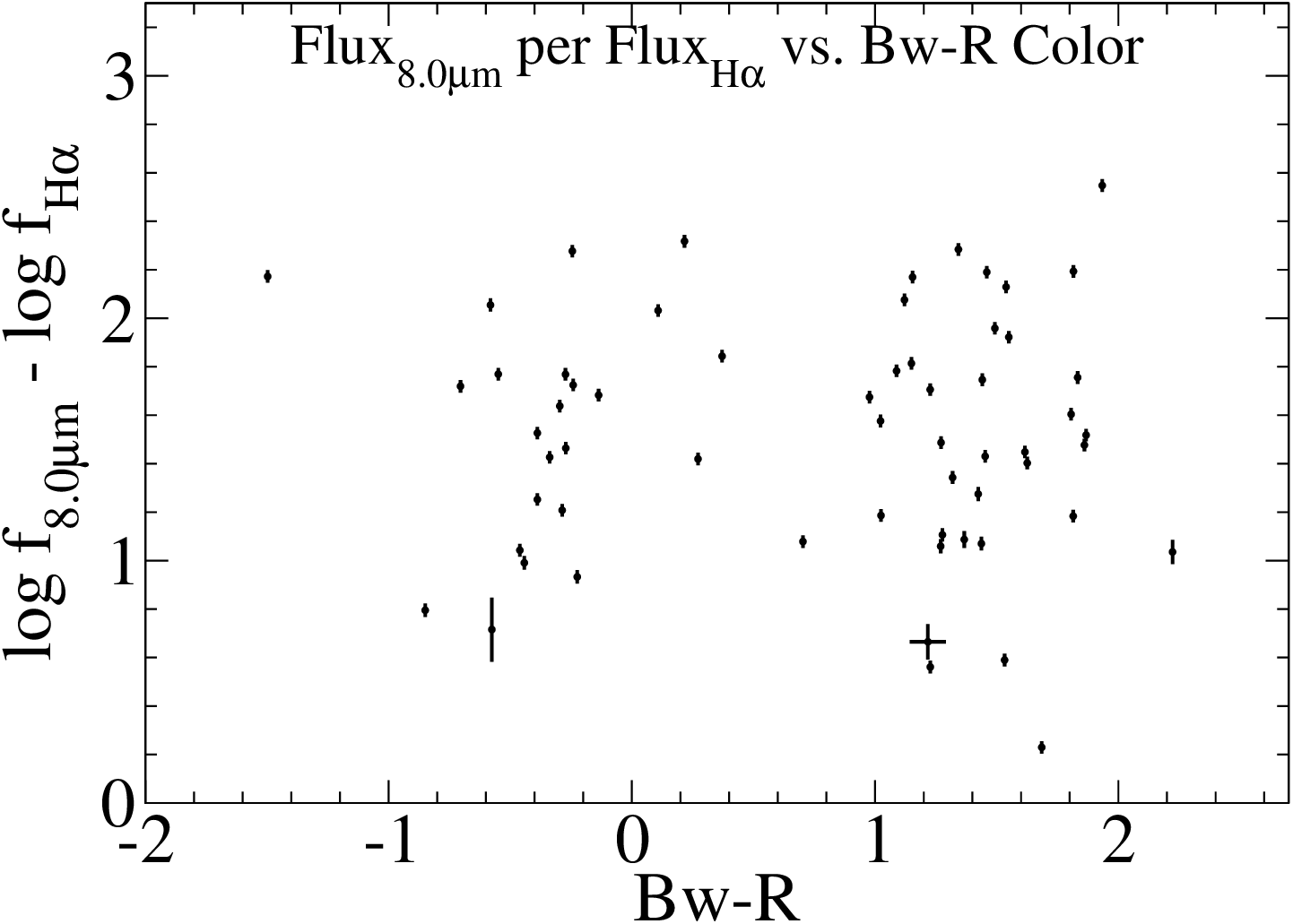}
  \caption{The ratio of 8.0\micron flux to extinction-corrected H$\alpha$ flux \vs Bw-R color; this figure shows no trend, with a Spearman rank-order coefficient of 0.02.}
  \label{fig:8micronefficiency_color}
\end{figure}


\begin{figure}
  \includegraphics[width=\linewidth]{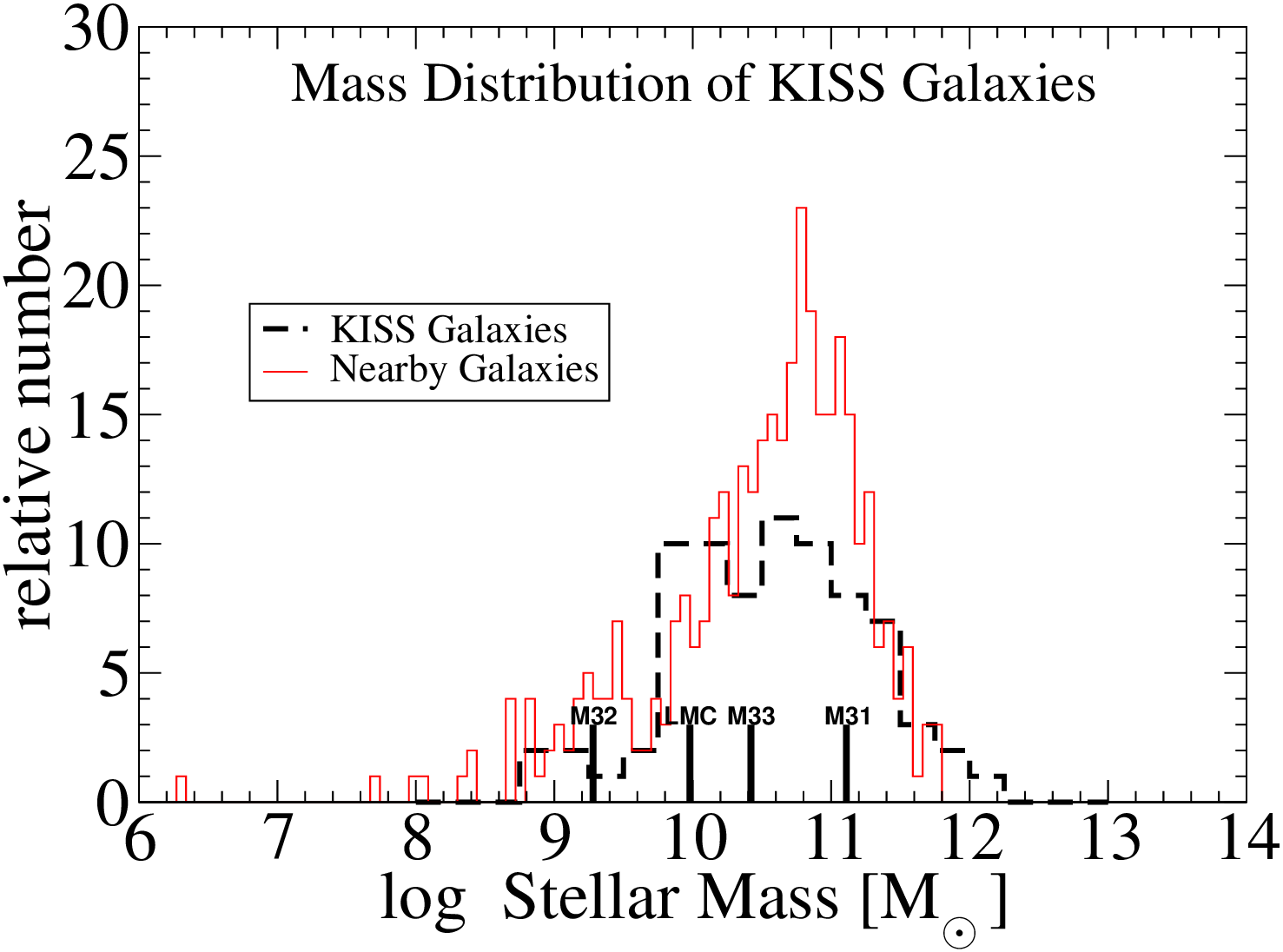}
  \caption{Distribution of stellar masses of the KISS galaxies (black) and Nearby Galaxies from \citep{GildePaz} (red); note that M31 falls near the upper end of the distribution.}
  \label{fig:masses}
\end{figure}

\begin{figure}
  \includegraphics[width=\linewidth]{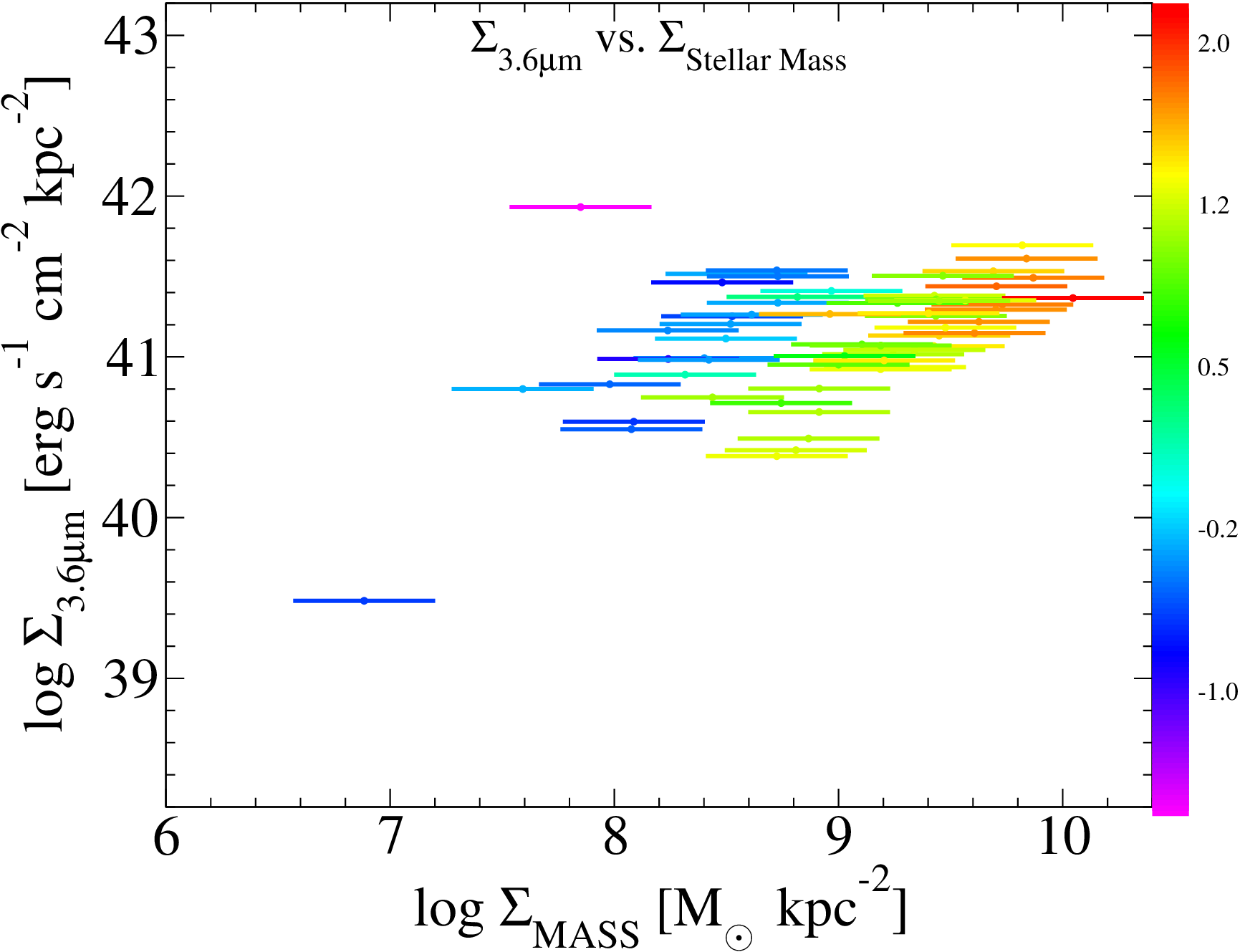}
  \caption{3.6\micron surface luminosity \vs to stellar mass surface density relationship.  Points are color-coded by Bw-R color with exact values indicated by the key on the right side of the plot. The horizontal (stellar mass) error bars are much larger, and are dominated by the 0.1 dex uncertainty in the mass-to-light ratio.  Note the upward trend, the large scatter, and the Bw-R gradient across the trend-line, indicating that 3.6\micron luminosity cannot be used as a stellar mass indicator without a color correction.}
  \label{fig:masstolight}
\end{figure}

\begin{figure}
  \includegraphics[width=\linewidth]{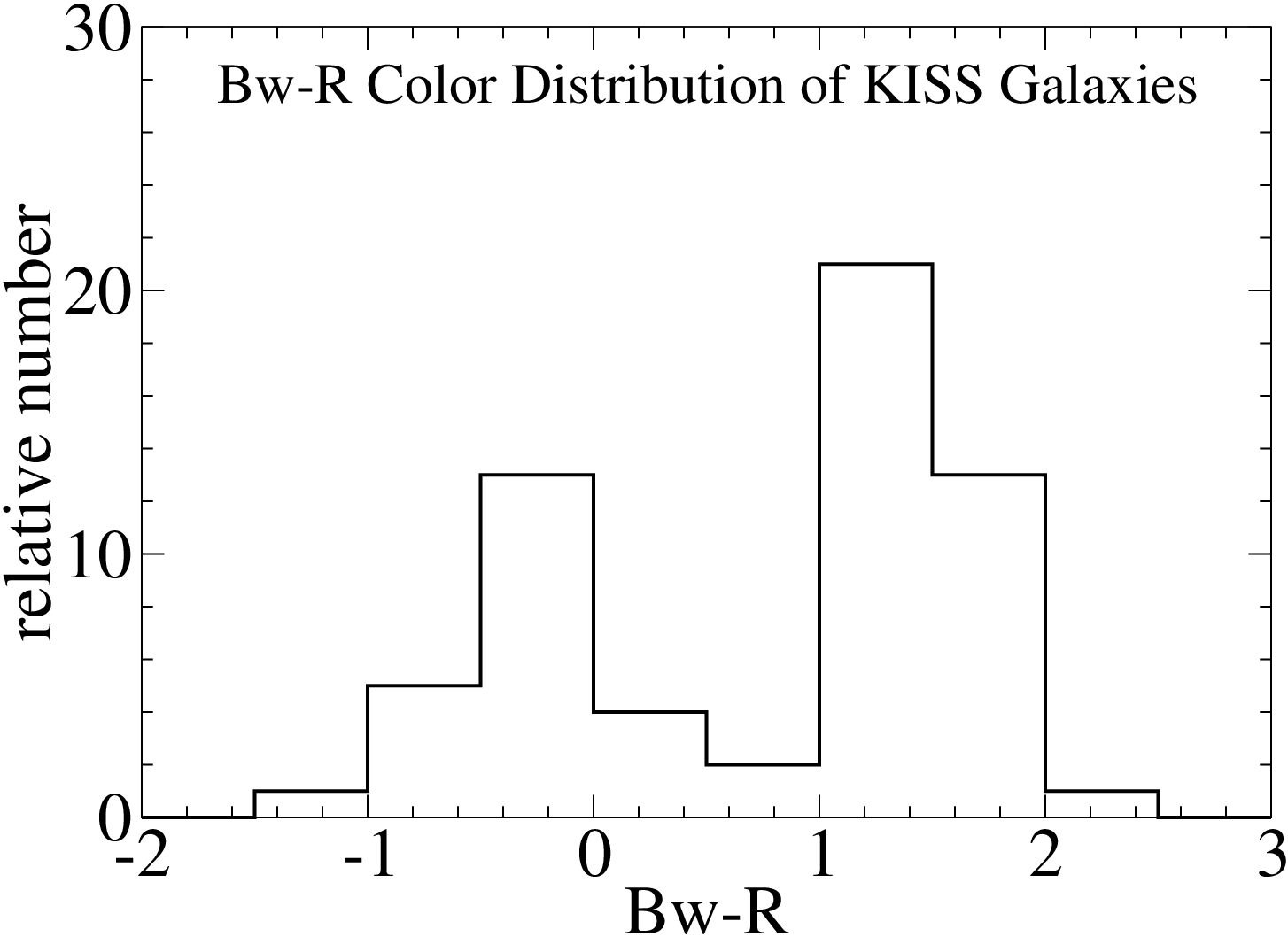}
  \caption{Histogram of Bw-R colors for the 76 KISS galaxies that have reliable optical colors.  Note that since the detection method used favors galaxy centers, the objects on the red side of the histogram are typically late-type galaxies with large bulges.}
  \label{fig:BwRhistogram}
\end{figure}

\begin{figure}
  \includegraphics[width=\linewidth]{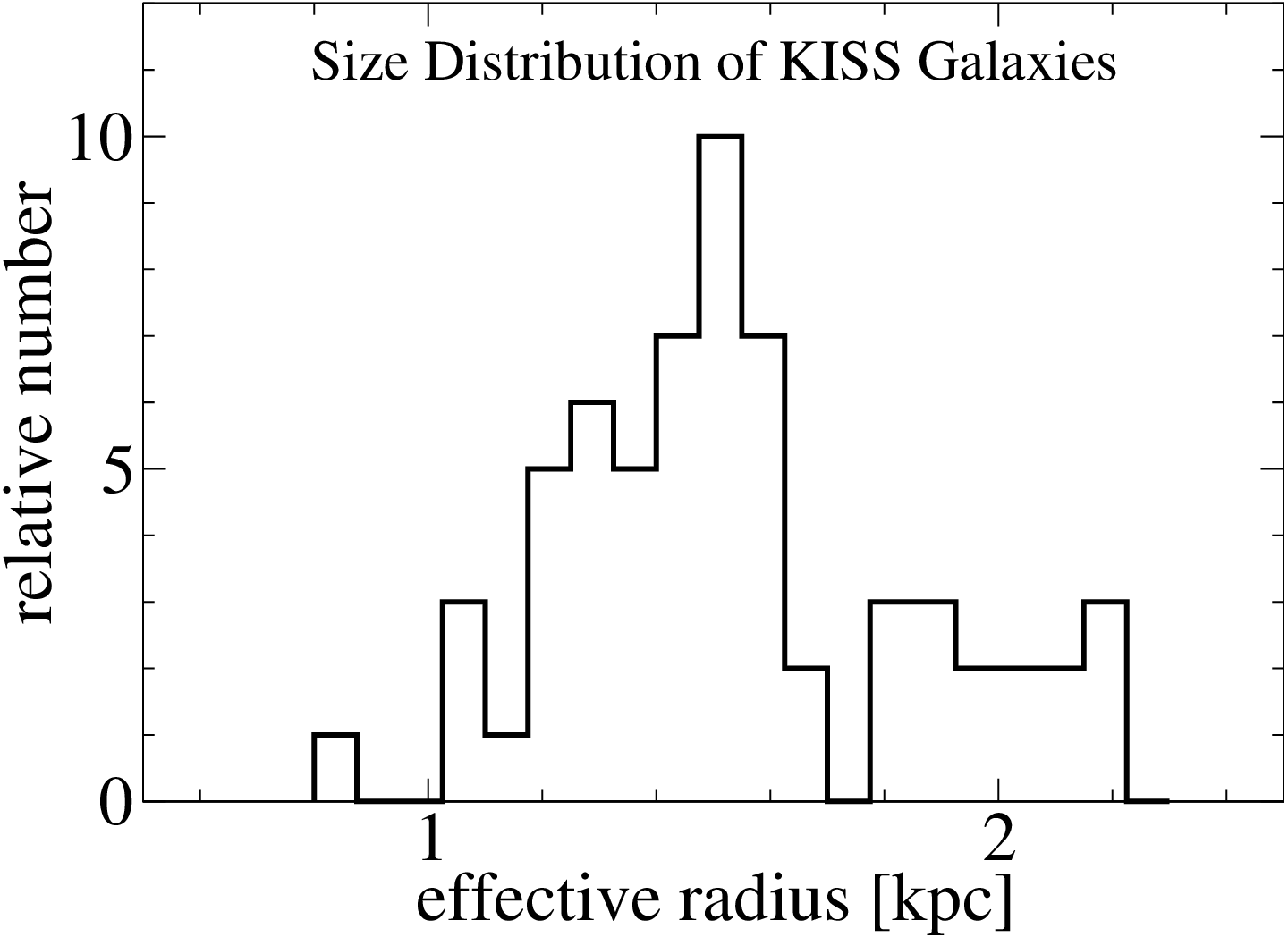}
  \caption{Distribution of effective radii of the KISS galaxies, taken as the geometric mean of the semi-major and semi-minor axes.  Note that the radii represented here correspond only to the areas of the galaxies that were detected as photometrically secure in IRAC 3.6\micron; the galactic disks likely extend beyond the radii shown here, but at surface brightnesses too low to allow direct comparison in a multi-wavelength analysis.}
  \label{fig:sizes}
\end{figure}